\documentclass[english,12pt]{iopart}
\usepackage[T1]{fontenc}
\usepackage[latin9]{inputenc}
\usepackage{geometry}
\usepackage{color}
\usepackage{babel}
\usepackage{array}
\usepackage{float}
\usepackage{booktabs}
\usepackage{textcomp}
\usepackage{multirow}
\usepackage{amstext}
\usepackage{amssymb}
\usepackage{graphicx}
\usepackage[numbers]{natbib}
\usepackage[unicode=true,
 bookmarks=true,bookmarksnumbered=false,bookmarksopen=false,
 breaklinks=false,pdfborder={0 0 0},pdfborderstyle={},backref=false,colorlinks=false]
 {hyperref}

\makeatletter

\providecommand{\tabularnewline}{\\}

\usepackage{iopams}
\usepackage{setstack}

\usepackage{amssymb}
\makeatother

\begin{document}
\title{Engineering the Electronic, Magnetic, and Optical Properties of GaP
Monolayer by Substitutional Doping: A first-principles study }
\author{Khushboo Dange, Rachana Yogi, and Alok Shukla{*}}
\address{Department of Physics, Indian Institute of Technology Bombay, Powai,
Mumbai 400076, India}
\ead{khushboodange@gmail.com, yogirachana04@gmail.com, shukla@iitb.ac.in}
\begin{abstract}
In this paper we present a thorough first-principles density functional
theory based computational study of the structural stability, electronic,
magnetic, and optical properties of pristine and doped gallium phosphide
(GaP) monolayers. The pristine GaP monolayer is found to have a periodically
buckled structure, with an indirect band gap of 2.15 eV. The doping
by X (B, Al, In, C, Si, Ge, Sn, Zn, and Cd) at the Ga site, and Y
(N, As, Sb, O, S, Se, Te, Zn, and Cd) at the P site is considered,
and an indirect to direct band gap transition is observed after doping
by In at the Ga site. For several cases, significant changes in the
band gap are seen after doping, while the system becomes metallic
when O is substituted at the P site. \textcolor{black}{The spin-polarized
band structures are calculated for the monolayers with doping-induced
magnetism, and we find that for some cases a direct band gap appears
for one of the spin orientations. For such cases, we investigate the
intriguing possibility of spin-dependent optical properties. Furthermore,
for several cases the band gap is very small for one of the spin orientations,
suggesting the possibility of engineering half metallicity by doping.
}For the layers with direct band gaps, the calculated optical absorption
spectra are found to span a wide energy range in the visible and ultraviolet
regions. \textcolor{black}{The computed formation energies of both
the pristine and doped structures are quite small, indicating that
the laboratory realization of such structures is quite feasible. On
the whole,} our results suggest that the doped GaP monolayer is a
material with potentially a wide range of applications in nanoelectronics,
spintronics, optoelectronics, solar cells, etc.
\end{abstract}
\maketitle
\textbf{\Large{}Keywords:} Gallium phosphide (GaP) monolayer; band
gap engineering; DFT; binding energy; formation energy; magnetism;
electronic and optical properties

\section{Introduction}

In the last couple of decades, with the rapid advances in technology,
nano-scaled systems have become the most intensively studied materials
\citep{baig2021,lalitha2019}. The successful synthesis of graphene
\citep{geim2004,geim2007} with its exclusive properties such as high
carrier mobility, good physical strength, 97.2 \% transparency, high
surface to volume ratio \citep{liu2007,shen2020,nair2008}, etc.,
have enthused the research in the field of 2D materials\textcolor{magenta}{{}
}\textcolor{black}{\citep{mas2011,zhang2016,long2019,bhimanapati2015,schaibley2016}.
}Synthesis of 2D materials by the techniques such as electrochemical
exfoliation \citep{liu2019ef,yang2019}, chemical vapor deposition
\citep{yu2015,zhang2020}, etc. is taking place at a rapid rate. 2D
materials synthesized till date possess unique electronic, optical,
chemical, magnetic, electrical and mechanical properties \citep{qiao2021,akinwande2017,thiel2019,yang2017,lee2018,mitta2020}.
They exhibit high carrier mobilities \citep{restrepo2014,mir2020},
good thermal conductivity \citep{wu2022} and high optical absorption
range \citep{wild2022,negri2019}. Such useful properties make them
potential candidates for wide range of applications including batteries
\citep{ye2020}, energy storage \citep{sahoo2016}, sensors \citep{anichini2018},
spintronics \citep{ahn2020,choudhuri2019}, photodetectors \citep{jiang2021},
optoelectronics \citep{bao2018,zhang2016van}, etc. 

In particular, binary 2D compounds made up of III-V group elements,
have attracted tremendous attention due to their interesting electronic
and optical properties, with potentially widespread applications \citep{csahin2009,zhuang2013,suzuki2015,lu2020,rouzhahong2020,liu2019str,nowak1994,zhao2021,wang2010}.
In spite of such useful properties and applications, 2D compounds
that have been synthesized from binary III-V group are limited \citep{singla2015,tsipas2013,koratkar2016,dai2022}.
Gallium nitride (GaN) monolayer has been synthesized \citep{koratkar2016}
and its properties were explored both experimentally and theoretically.
GaN monolayer possesses an indirect band gap of around 2 eV, realized
by density functional theory (DFT) calculations \citep{li2021}, which
impede its potential for optoelectronic devices \citep{al2016}. Remarkable
efforts have been made to undergo indirect to direct transition by
using the strategy of doping \citep{chen2016,xia2013}, adsorption
\citep{chen2019,cui2020}, functionalization \citep{du2021,baur2005},
applying electric field \citep{xu2013} and strain \citep{shu2019,bahuguna2018}.
Similarly, 2D aluminium nitride (AlN) has been synthesized \citep{tsipas2013}
and studied by scientific community \citep{peng2013,beshkova2020}.
\textcolor{black}{However, some of the III-V monolayers which are
predicted \citep{csahin2009,zhuang2013,suzuki2015} to have potential
for transformative electronic and optical properties are awaiting
discovery. }

Bulk Gallium Phosphide (GaP) is an important semiconductor with a
wide indirect band gap of 2.26 eV, grown in crystalline form using
epitaxial techniques \citep{aparna2014,treece1992}. GaP exhibits
interesting properties such as high refractive index, good mechanical
durability, spectral transmission range in the infrared region, etc.
\citep{vaclavik2013}. These set of properties make GaP a preferable
candidate for designing optical elements such as fast lenses, and
it opens a route towards visible and near infrared applications\textcolor{black}{{}
\citep{wight1977,vaclavik2013}.} It also plays essential role in
high-temperature transistors \citep{zipperian1981}, photovoltaic
solar cells \citep{lu2012}, optical devices \citep{singh2013}, etc.
However in bulk form, modification of electronic and optical properties
of any material is possible only up to a certain extent, which can
be enhanced further in its reduced dimensional form. As a result,
it is important to theoretically study the electronic structure, optical,
and magnetic properties of pristine and doped GaP monolayer using
a first-principles approach so as to provide some guidance to the
experimentalists. GaP thin films were also fabricated using different
techniques \citep{emmer2017,mori1987,tilmann2020} and were found
to have honeycomb like structure similar to silicene \citep{ren2016}.
Hexagonal GaP monolayer with slightly buckled honeycomb like structure
is predicted to have wide indirect band gap of 2.96 eV computed using
HSE06 functional \citep{suzuki2015}, which is higher compared to
its bulk counterpart due to quantum confinement effect. Sahin et al.
have studied the geometry of III-V monolayers including GaP, along
with their electronic and mechanical properties \citep{csahin2009}.
The GaP monolayer is predicted to be stable in the buckeld form \citep{csahin2009}.
The dynamic stability of hexagonal GaP monolayer has already been
confirmed by the absence of imaginary phonon frequencies \citep{csahin2009,zhuang2013,suzuki2015}.
Zhuang et al. \citep{zhuang2013} used hybrid density functional and
the $G_{0}W_{0}$ method to determine the stability and electronic
structure of III-V monolayers including GaP. They also demonstrated
that GaP monolayer can be synthesized on lattice-matched substrates.
Suzuki in 2015 also reported that III-V monolayers are the better
candidates for semiconductor devices \citep{suzuki2015}. A detailed
study on the electronic properties of hexagonal GaP monolayer has
been reported by Kumar et al. \citep{kumar2016} in 2016. Akbari et
al. studied the effect of biaxial strain on AlP and GaP monolayers
\citep{akbari2018}, and showed that it is a viable way to tune their
electronic as well as optical properties. Facilitating the indirect
to direct band gap transition of GaP monolayer using any of the viable
way such as doping, functionalization, etc., could be beneficial for
its application in the field of optoelectronic devices. \textcolor{black}{Various
techniques have been used to tune the electronic properties of nanomaterials
such as: passivation \citep{ma2018}, adsorption \citep{liu2019},
doping \citep{zhang2000}, defects formation \citep{hahn1999}, termination
\citep{hart2019}, functionalization \citep{liu2015} etc. Because
doping is much easier and inexpensive to achieve in a laboratory,
it is believed to be one of the best ways to tailor the electronic
properties of nanomaterials}. To the best of our knowledge, the effects
of substitutional doping on GaP monolayer have not been reported in
the literature till date. Therefore, in this work, we study the effects
of substitutional doping on the electronic, magnetic and optical properties
of GaP monolayer using a first-principles DFT based methodology. 

In this work, we perform a systematic study of the influence of substitutional
doping by the same group elements, those from the nearest neighboring
groups, and transition metals on the properties of GaP monolayer.
In particular, we have studied the structural stability, electronic
structure, magnetic, and optical properties of pristine and eighteen
substitutionally doped GaP monolayers. We show that for certain dopants,
the band gap of GaP monolayer can be tuned from indirect to direct,
semiconductor to metallic, and semiconductor to half-metallic. The
magnetic behavior is also observed in certain cases, which is important
for spin-dependent transport and optical applications. The variation
in Fermi energy also suggests various doping candidates for GaP monolayer.
Optical properties of these substitutionally doped GaP monolayer were
also investigated, and for several cases spin-dependent optical response
was confirmed.

The remainder of this paper is organized as follows. In the next section
we briefly discuss our computational approach. Next, in section \ref{sec:Results-and-Discussion}
we present and discuss our results. Finally, in section \ref{sec:Conclusion}
we present our conclusions and also discuss future prospects.

\section{Computational Details}

All the calculations were performed within the framework of first-principles
based DFT \citep{hohenberg1964,kohn1965} using plane wave basis set
as implemented in Vienna Ab initio Simulation Package (VASP)\citep{kresse1996,Georg1996}.
Perdew-Burke-Ernzerhof (PBE) exchange-correlation functional within
the generalized gradient approximation (GGA) \citep{perdew1996} along
with projected augmented wave (PAW) potential \citep{kresse1999,blochl1994}
was used to solve the Kohn-Sham equations self-consistently.\textcolor{black}{{}
In order to understand the convergence of our calculations with respect
to the size of the supercell, we considered three supercells of the
GaP monolayer of increasing sizes, namely $2\times2$, $4\times4$,
and $8\times8$, consisting of 8, 32, and 128 atoms, respectively,
as discussed in the next section. Subsequently, to balance off the
requirements of numerical precision with the computational effort,
we performed the rest of our calculations employing the $4\times4$
supercell (see Fig. \ref{Fig:1}) both for the pristine as well as
doped systems.} For the doped systems, a single Ga or P atom was substituted
by the dopant atom. For the Brillouin zone sampling, Monkhorst-Pack
\citep{monkhorst1976} k-point grids of 5$\times$5$\times$1 and
7$\times$7$\times$1 were taken for geometry optimization, and self-consistent
field (SCF) calculations, respectively. A vacuum space of 12 Å along
the z-direction was provided to avoid the interaction between two
consecutive layers. The cutoff energy of 500 eV was set for plane
wave basis set. The convergence criteria of $10^{-5}$ eV was set
to reach the self-consistency. The geomtery optimization iterations
were continued until the atomic forces on each atom were less than
0.01 eV/Å.\textcolor{black}{{} Further, a more accurate exchange-correlation
functional, HSE06 \citep{heyd2003hybrid}, is also employed after
GGA-PBE \citep{perdew1996} to repeat some of the calculations. All
the geometry relaxations and SCF calculations were performed within
the spin-polarized formalism. Ab initio molecular dynamics (AIMD)
\citep{Kresse1994aimd} simulations are also performed for the doped
systems to confirm their stability at high temperatures. For magnetically
polarized monolayers, the majority electrons are assumed to have up
spin, while minority ones ha}ve down spin. For those dopants which
were not strongly bound to the monolayer, we also accounted for the
van der Waal interactions by including the DFT-D3 \citep{grimme2010}.
In order to obtain a deeper understanding of the nature of bonding
in the systems, we also performed the Bader charge analysis using
the code developed by Henkelman group\citep{tang2009}.

\section{Results and Discussion}

\label{sec:Results-and-Discussion}

\subsection{\textcolor{black}{Convergence with respect to the supercell size}}

\textcolor{black}{The results of our calculations on the pristine
monolayer considering supercells of the sizes $2\times2$, $4\times4$,
and $8\times8$, are presented in Table \ref{tab:Convergence}. From
the table it is obvious that there is perfect agreement among the
results obtained using three supercells as far as the geometric parameters
such as bond lengths and bond angles are concerned. However, we note
small disagreements in the values of: (a) total energy per atom ($\approx0.2\%$),
and (b) binding energy per atom ($\approx0.26\%$). Because such small
disagreements are quite acceptable from a computational point of view,
we decided to perform the remainder of our calculations, both on the
pristine and doped systems using a $4\times4$ supercell. }

\textcolor{black}{}
\begin{table}
\textcolor{black}{\caption{Convergence of various parameters with respect to the supercell size.\label{tab:Convergence}}
}
\centering{}\textcolor{black}{}%
\begin{tabular}{cccc}
\toprule 
\textcolor{black}{Supercell} & \textcolor{black}{$2\times2$} & \textcolor{black}{$4\times4$} & \textcolor{black}{$8\times8$}\tabularnewline
\midrule 
\textcolor{black}{Number of atoms} & \textcolor{black}{8} & \textcolor{black}{32} & \textcolor{black}{128}\tabularnewline
\textcolor{black}{Lattice constant ($\text{Å}$)} & \textcolor{black}{3.85} & \textcolor{black}{3.85} & \textcolor{black}{3.85}\tabularnewline
\textcolor{black}{Ga-P bond length ($\text{Å}$)} & \textcolor{black}{2.27} & \textcolor{black}{2.27} & \textcolor{black}{2.27}\tabularnewline
\textcolor{black}{GaPGa bond angle (degrees)} & \textcolor{black}{115.575} & \textcolor{black}{115.575} & \textcolor{black}{115.575}\tabularnewline
\textcolor{black}{PGaP bond angle (degrees)} & \textcolor{black}{115.573} & \textcolor{black}{115.573} & \textcolor{black}{115.573}\tabularnewline
\textcolor{black}{Total Energy per atom (eV)} & \textcolor{black}{-4.097} & \textcolor{black}{-4.098} & \textcolor{black}{-4.106}\tabularnewline
\textcolor{black}{Binding Energy per atom(eV)} & \textcolor{black}{-3.045} & \textcolor{black}{-3.045} & \textcolor{black}{-3.053}\tabularnewline
\bottomrule
\end{tabular}
\end{table}

\subsection{Structure and Stability}

\begin{figure}[H]
\begin{centering}
\includegraphics[scale=0.45]{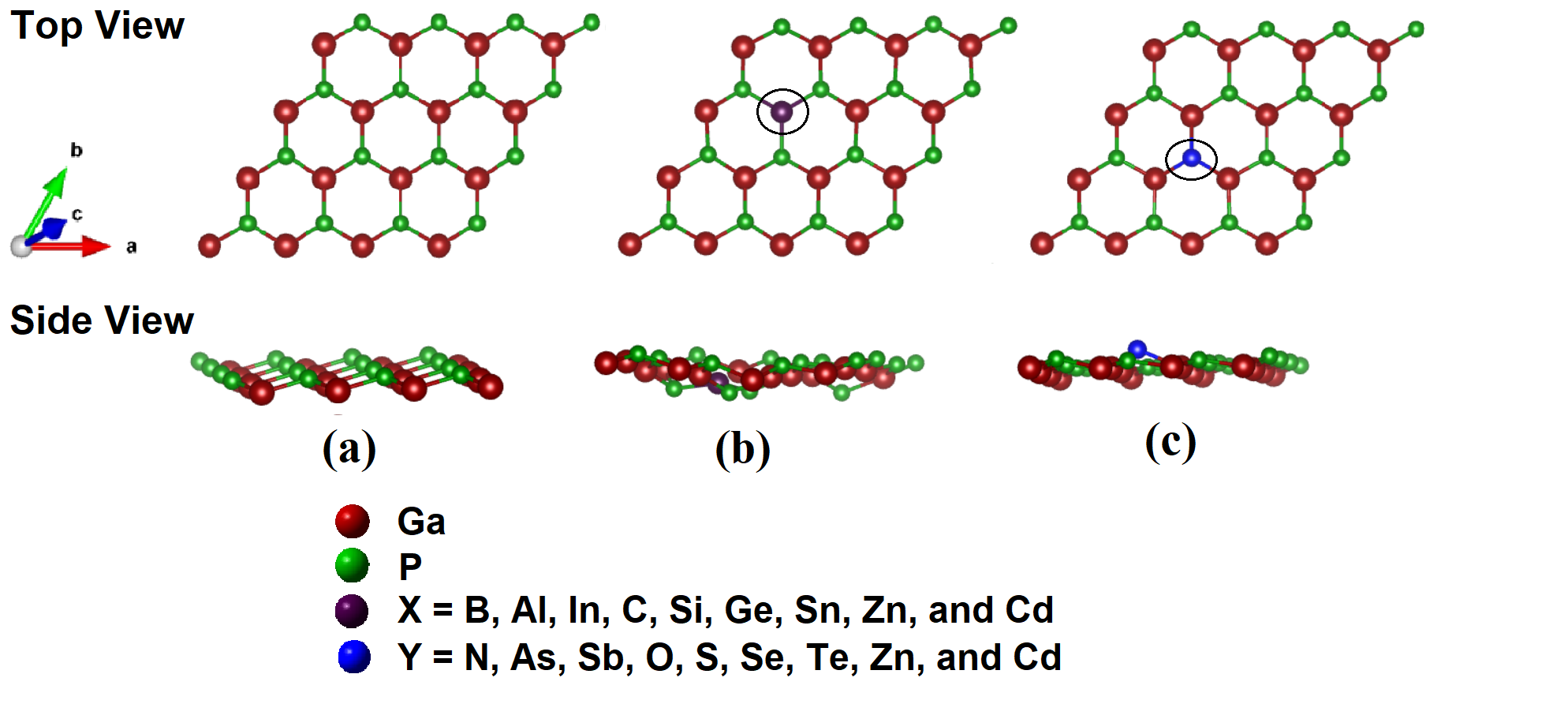}
\par\end{centering}
\caption{\label{Fig:1}Optimized structures of the: (a) pristine GaP monolayer,
substitutionally doped GaP monolayers with the dopant at the; (b)
Ga site, and (c) P site. }
\end{figure}

\textcolor{black}{The optimized structure of pristine GaP monolayer
is shown in Fig.\ref{Fig:1}(a) which is a periodically buckled honeycomb
like structure similar to that of silicene \citep{ren2016}. In the
pristine GaP monolayer, each Ga(P) atom is bonded to three nearest-neighbor
P(Ga) atoms. Thus, due to the translational symmetry, all Ga(P) atoms
are chemically equivalent to other Ga(P) atoms. We note that for all
the three supercells of the the pristine monolayer (see Table \ref{tab:Convergence}),
the optimized lattice constant of 3.85 Å, and the bond angles 115.575$\text{°}$
($\angle$GaPGa), 115.573$\text{°}$ ($\angle$PGaP) are in close
agreement with the previously reported values \citep{csahin2009,zhuang2013,akbari2018}.
Furthermore, the optimized Ga-P bond length 2.27 Å is also in good
agreement with the values reported in the literature\citep{csahin2009,kumar2016}.}\textcolor{blue}{{} }

We also present the optimized bond lengths and bond angles of the
doped structures \textcolor{black}{computed using the $4\times4$
supercell} in Table \ref{tab1}. We note that after doping, the Ga-P
bond length varies in the range 2.27--2.41 Å, which signals an average
increase. The bond angles $\angle$GaPGa and $\angle$PGaP undergo
more significant changes on doping as compared to the pristine case,
with the values in the ranges 102.85$\text{°}$--128.15$\text{°}$,
and 96.95$\text{°}$--128.60$\text{°}$, respectively. Therefore,
it is quite obvious that doping by a single atom has a very significant
impact on the structural parameters of the monolayer.

\begin{table}[H]
\centering{}\caption{\label{tab1}The optimized bond lengths and bond angles for the doped
structures under investigation.}
\begin{tabular}{cccc}
\toprule 
\multicolumn{2}{c}{Bond Length (Å)} & \multicolumn{2}{c}{Bond Angle ( degrees)}\tabularnewline
\midrule 
Ga(X)P & GaP(Y) & Ga(X)P & GaP(Y)\tabularnewline
\midrule 
Ga-P=2.28-2.37 & Ga-P=2.27-2.41 & $\angle GaPGa$= 102.85-126.68 & $\angle GaPGa$= 103.24-128.15\tabularnewline
P-B = 1.93 & Ga-N=1.94 & $\angle PGaP$=106.23-124.45 & $\angle PGaP$=96.95-128.60\tabularnewline
P-Al=2.33 & Ga-As=2.39 & $\angle PBP$=117.96 & $\angle GaNGa$=119.97\tabularnewline
P-In=2.47 & Ga-O=2.0 & $\angle PAlP$=117.33-117.35 & $\angle GaAsGa$=104.53\tabularnewline
P-C=1.83 & Ga-S=2.31-2.6 & $\angle PInP$=115.10 & $\angle GaOGa$=119.97\tabularnewline
P-Si=2.28 & Ga-Se=2.44-2.68 & $\angle PCP$=117.50 & $\angle GaSGa=$117.69-119.29\tabularnewline
P-Sn=2.55 &  & $\angle PSiP$=103.29 & $\angle GaSeGa$=113.23-115.13\tabularnewline
P-Zn=2.31-2.33 &  & $\angle PSnP$=94.99 & \tabularnewline
P-Cd=2.53 &  & $\angle PZnP$=112.07-123.37 & \tabularnewline
 &  & $\angle PCdP$=116.46-121.75 & \tabularnewline
\bottomrule
\end{tabular}
\end{table}

To ensure the thermal stability of the pristine structure, the binding
energy per atom ($E_{b}$) has been calculated using the relation

\begin{equation}
E_{b}=\frac{1}{N}\left[E_{GaP}-(N_{Ga}E_{Ga}+N_{P}E_{P})\right]\label{eq:binding-energy}
\end{equation}

where $E_{GaP}$ is the total energy of pristine GaP monolayer, $E_{Ga}$,
and $E_{P}$ are the energies of isolated Ga and P atoms, respectively.
Further, $N_{Ga}$ and $N_{P}$ denote the numbers of Ga and P atoms,
respectively in the supercell, while $N=N_{Ga}+N_{P}$ is the total
number of atoms. The resultant binding energy calculated using Eq.
\ref{eq:binding-energy} is -3.045 eV/atom which confirms that the
structure is thermodynamically stable. GaP monolayer is energetically
less stable than its 3D bulk counterpart, with a relative energy of
0.45 eV/atom \citep{zhuang2013}. However, it is less than the relative
energy of the 2D hexagonal monolayer of SiC (0.50 eV/atom) which has
recently been synthesized \citep{sic-synthesis-2021}. \textcolor{black}{Another
important quantity from an experimental point of view is the formation
energy $(E_{f}$) of the pristine GaP monolayer defined as\citep{Ullah2020}}

\textcolor{black}{
\begin{equation}
E_{f}=\left[E_{GaP}-(N_{Ga}\mu_{Ga}+N_{P}\mu_{P})\right],\label{eq:formation}
\end{equation}
}

\textcolor{black}{where $\mu_{Ga}$ and $\mu_{P}$, respectively,
represent the chemical potentials of Ga and P corresponding to their
most common bulk phases. We performed DFT calculations of the chemical
potentials of various elements using the VASP code\citep{Georg1996,kresse1996,kresse1999},
and the computed values are presented in Table S1 of the SI (supporting
information). Employing those values of chemical potentials, the computed
value of the formation energy of the pristine GaP monolayer is rather
small at 0.032 eV, indicating that the synthesis of GaP monolayer
is chemically feasible. Furthermore, the dynamical stability of the
GaP monolayer has been verified by several authors in previous works
by computing its phonon spectra, in which no imaginary frequencies
were found \citep{csahin2009,zhuang2013,suzuki2015}. }

Next, we optimized the geometries of the substitutionally doped GaP
monolayers, and the results are presented in Fig.\ref{Fig:1}(b),(c).
We denote the doped structures as Ga(X)P and GaP(Y), where X(=B, Al,
In, C, Si, Ge, Sn, Zn and Cd) and Y(=N, As, Sb, O, S, Se, Te, Zn,
and Cd) are the dopants replacing Ga and P atoms, respectively. Of
all the optimized structures obtained by substituting the Ga atom,
we note that the Ge-doped structure results in the physisorption of
the dopant atom, whereas all other dopants are involved in chemical
bond formation with their nearest neighbor host atoms. Similarly,
in case of the substitution of P atom, we note that Sb, Te, Zn, and
Cd doping resulted in physisorption, whereas other dopants are involved
in chemical bond formation with the host atoms. For the doped GaP
monolayers which exhibit physisorption, we have performed DFT-D3 calculations
using Grimme dispersion correction method \citep{grimme2010} to account
for the van der Waal contribution to the binding energy. However,
we found that in all the cases of physisorption, the contribution
of the Grimme correction to the total energy was insignificant (less
than $1.0\times10^{-6}$ eV), except for Ga(Ge)P for which it was
5.5 meV and has been included in the binding energy reported in Table
\ref{tab2}. To verify the structural stability, first we have calculated
the binding energies of all the optimized structures obtained by doping
impurity elements in GaP monolayer. The binding energies per atom
for the doped systems, $E_{b}^{(d)}$, are calculated by using the
expression 

\begin{equation}
E_{b}^{(d)}=\frac{1}{N}\left[E_{tot}-(N_{Ga}E_{Ga}+N_{P}E_{P}+E_{X/Y})\right]
\end{equation}

where $E_{tot}$ is the total energy of doped GaP monolayer, $E_{Ga}$,
$E_{P}$, and $E_{X/Y}$ respectively, represent the total energies
of the isolated Ga, P, and dopant atoms, and $N_{Ga}$, $N_{P}$,
denotes the number of Ga, P atoms in the doped system, respectively,
with $N=N_{Ga}+N_{P}+1$. The obtained values of $E_{b}^{(d)}$ along
with the binding energy relative to that of the pristine monolayer,
$\Delta E_{b}^{(d)}=E_{b}^{(d)}-E_{b}$ for Ga and P substituted cases
are presented in Tables \ref{tab2} and \ref{tab3}, respectively.
We note that the binding energy per atom for the doped systems in
most of the cases is lower than that of the pristine system, as also
indicated by the negative sign of the relative binding energy $\Delta E_{b}^{(d)}$
in those cases, suggesting higher stability as compared to the pristine
monolayer. The magnitudes of the relative binding energies $\Delta E_{b}^{(d)}$
for these systems are in the range 5 meV -- 140 meV, indicating their
thermodynamic stability and practical realization. However, opposite
is the case for doping with the transition metals Zn and Cd in place
of both Ga and P, \textcolor{black}{and also for doping with As and
Te in place of P}\textcolor{blue}{{} }for which $\Delta E_{b}^{(d)}$>0,
indicating instability of the lattice, possibly due to the large ionic
radii of the dopant atoms. This suggests that for these cases the
resulting lattice will be highly \textcolor{black}{strained. We also
computed the formation energies for the doped systems, $E_{f}^{(d)}$,
using the expression \citep{Hamid2022}}

\textcolor{black}{
\begin{equation}
E_{f}^{(d)}=\left[E_{tot}-(E_{GaP}-\mu_{Ga/P}+\mu_{X/Y})\right],\label{eq:doped_formation}
\end{equation}
}

\textcolor{black}{where $\mu_{Ga/P}$ and $\mu_{X/Y}$ denotes the
chemical potential of a Ga/P atom removed and considered doping element
X/Y, respectively. The resultant formation energies ($E_{f}^{(d)}$)
computed using the values of the chemical potentials listed in  Table
S1 of the SI are presented in Tables \ref{tab2} and \ref{tab3} for
the Ga and P substituted cases, respectively. The negative formation
energies obtained for all the doped cases except Ga(C)P, GaP(As),
GaP(Zn), and GaP(Cd) systems suggests that the formation of these
substitutional doped GaP monolayer is practically feasible, and the
doped monolayers will be more stable as compared to their pristine
counterpart. Further, the small positive values of $E_{f}^{(d)}$
for Ga(C)P, GaP(As), GaP(Zn), and GaP(Cd) systems are favorable according
to the literature \citep{Hamid2023} and thus the practical realization
of these doped GaP layers is also possible. }

\begin{table}[H]
\begin{centering}
\caption{\label{tab2}The calculated binding energy ($E_{b}^{(d)}$), relative
binding energy ($\Delta E_{b}^{(d)}$), formation energy ($E_{f}^{(d)})$,
magnetic moment ($\mu$, in Bohr magnetons), Fermi energy ($E_{F}$),
and electronic band gap ($E_{g})$ of all the structures obtained
by the substitutional doping of pristine GaP monolayer in place of
Ga atom (Ga(X)P). }
\begin{tabular}{cccccccc}
\toprule 
\multirow{2}{*}{Configurations} & \multirow{2}{*}{$E_{b}^{(d)}$ (eV)} & \multirow{2}{*}{$\Delta E_{b}^{(d)}$ (eV)} & \multirow{2}{*}{$E_{f}^{(d)}$(eV)} & \multirow{2}{*}{$\mu$} & \multirow{2}{*}{$E_{F}$(eV)} & \multicolumn{2}{c}{$E_{g}$(eV)}\tabularnewline
\cmidrule{7-8} \cmidrule{8-8} 
 &  &  &  &  &  & majority spin  & minority spin \tabularnewline
\midrule 
GaP (pristine) & -3.045 & 0.000 & \textcolor{black}{0.032} & 0.00 & -3.06 & 2.15 I & \tabularnewline
Ga(B)P & -3.185 & -0.140 & \textcolor{black}{-0.782} & 0.00 & -3.19 & 1.81 I & \tabularnewline
Ga(Al)P & -3.116 & -0.071 & \textcolor{black}{-1.451} & 0.00 & -3.14 & 2.00 I & \tabularnewline
Ga(In)P & -3.050 & -0.005 & \textcolor{black}{-0.576} & 0.00 & -3.12 & 2.00 D & \tabularnewline
Ga(C)P & -3.164 & -0.119 & \textcolor{black}{1.437} & 0.52 & -2.00 & 0.64 I & 1.25 I\tabularnewline
Ga(Si)P & -3.117 & -0.072 & \textcolor{black}{-0.402} & 0.42 & -2.26 & 1.21 I & 1.57 D\tabularnewline
Ga(Ge)P & -3.095 & -0.050 & \textcolor{black}{-0.432} & 0.49 & -2.52 & 1.45 I & 1.11 D\tabularnewline
Ga(Sn)P & -3.072 & -0.027 & \textcolor{black}{-0.424} & 0.55 & -2.65 & 1.50 I & 0.85 D\tabularnewline
Ga(Zn)P & -3.016 & 0.029 & \textcolor{black}{-0.685} & 0.44 & -3.24 & 2.21 I & 0.21 D\tabularnewline
Ga(Cd)P & -2.995 & 0.050 & \textcolor{black}{-0.416} & 0.46 & -3.26 & 2.12 D  & 0.12 D\tabularnewline
\bottomrule
\end{tabular}
\par\end{centering}
{\small{}I $\equiv$ Indirect gap and D $\equiv$ Direct gap}{\small\par}
\end{table}

\begin{table}[H]
\caption{\label{tab3}The calculated binding energy ($E_{b}^{(d)}$), relative
binding energy ($\Delta E_{b}^{(d)}$), formation energy ($E_{f}^{(d)})$,
magnetic moment ($\mu$, in Bohr magnetons), Fermi energy ($E_{F}$),
and electronic band gap ($E_{g})$ of all the structures obtained
by the substitutional doping of pristine GaP monolayer in place of
P atom (GaP(Y)).}

\begin{centering}
\begin{tabular}{cccccccc}
\toprule 
\multirow{2}{*}{Configurations} & \multirow{2}{*}{$E_{b}^{(d)}$(eV)} & \multirow{2}{*}{$\Delta E_{b}^{(d)}$ (eV)} & \multirow{2}{*}{$E_{f}^{(d)}$(eV)} & \multirow{2}{*}{$\mu$} & \multirow{2}{*}{$E_{F}$(eV)} & \multicolumn{2}{c}{$E_{g}$(eV)}\tabularnewline
\cmidrule{7-8} \cmidrule{8-8} 
 &  &  &  &  &  & majority spin & minority spin\tabularnewline
\midrule 
GaP(N) & -3.131 & -0.086 & \textcolor{black}{-1.027} & 0.00 & -3.14 & 1.64 I & \tabularnewline
GaP(As) & -3.023 & 0.022 & \textcolor{black}{0.912} & 0.00 & -3.06 & 2.07 I & \tabularnewline
GaP(Sb) & -3.059 & -0.014 & \textcolor{black}{-1.353} & 0.00 & -3.26 & 2.13 I & \tabularnewline
GaP(O) & -3.130 & -0.085 & \textcolor{black}{-2.828} & 0.00 & -3.19 & M & \tabularnewline
GaP(S) & -3.085 & -0.040 & \textcolor{black}{-1.545} & 0.54 & -2.01 & 0.99 I & 1.71 D\tabularnewline
GaP(Se) & -3.057 & -0.012 & \textcolor{black}{-1.172} & 0.55 & -2.52 & 0.88 I & 1.21 D\tabularnewline
GaP(Te) & -3.029 & 0.016 & \textcolor{black}{-0.492} & 0.54 & -3.12 & 1.09 I & 1.44 D\tabularnewline
GaP(Zn) & -2.941 & 0.104 & \textcolor{black}{0.934} & 0.60 & -2.58 & 0.96 D & 0.23 I\tabularnewline
GaP(Cd) & -2.938 & 0.107 & \textcolor{black}{0.606} & 0.55 & -2.51 & 0.88 D & 0.13 I\tabularnewline
\bottomrule
\end{tabular}
\par\end{centering}
{\small{}I$\equiv$Indirect, M$\equiv$Metallic, and D $\equiv$ Direct}{\small\par}
\end{table}

\begin{figure}
\begin{centering}
\includegraphics[scale=0.5]{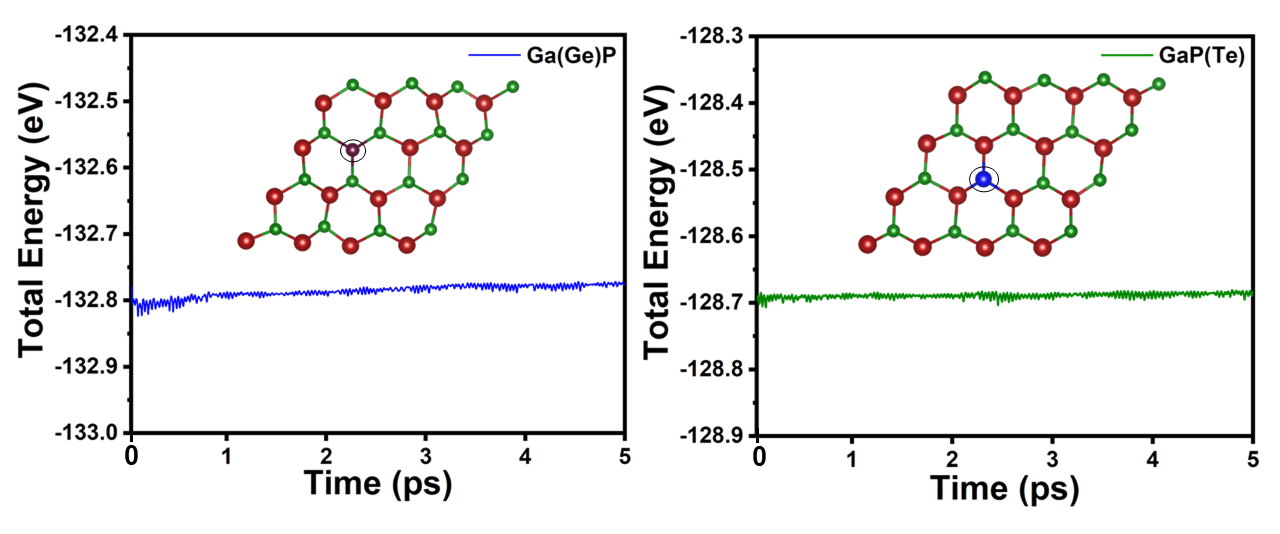}
\par\end{centering}
\caption{\label{fig:md}\textcolor{black}{The variations of the total energy
of (a) Ga(Ge)P and (b) GaP(Te) monolayers with time at 500K. Corresponding
geometries are shown in inset.}}
\end{figure}

\textcolor{black}{Further, we have also performed the AIMD \citep{Kresse1994aimd}
calculations for all the doped GaP monolayers. These simulations are
useful for validating the stability of systems at high temperatures
\citep{waheed2023,waheed2023janus,idress2023}, and were performed
in the canonical ensemble for 5 ps with a time step of 1 fs at a constant
temperature of 500K by using the Nose-Hoover thermostat \citep{nose1984}.
The total energy of all the doped systems is plotted as a function
of time steps for all the considered systems (see SI), while the results
for Ga(Ge)P and GaP(Te) systems are presented in Fig. \ref{fig:md}.
We note that the total average energy remains constant in spite of
slight variation in energies with time. The corresponding geometries
remain stable at 500K as shown in the inset. The changes in Ga-P bond
lengths are 0.09 Å and 0.06 Å for the Ga(Ge)P and GaP(Te) systems,
respectively, whereas the change in P-Ge and Ga-Te bond lengths are
0.07 Å and 0.007 Å, respectively. The results of AIMD simulations
for the rest of the doped cases show similar behavior as depicted
in Figs. S1 and S2 of the SI. These findings suggests that all the
doped GaP monolayers are thermodynamically stable at high temperatures.}

\subsection{Electronic Structure}

\subsubsection{Pristine Monolayer}

\begin{figure}
\begin{centering}
\includegraphics[scale=0.4]{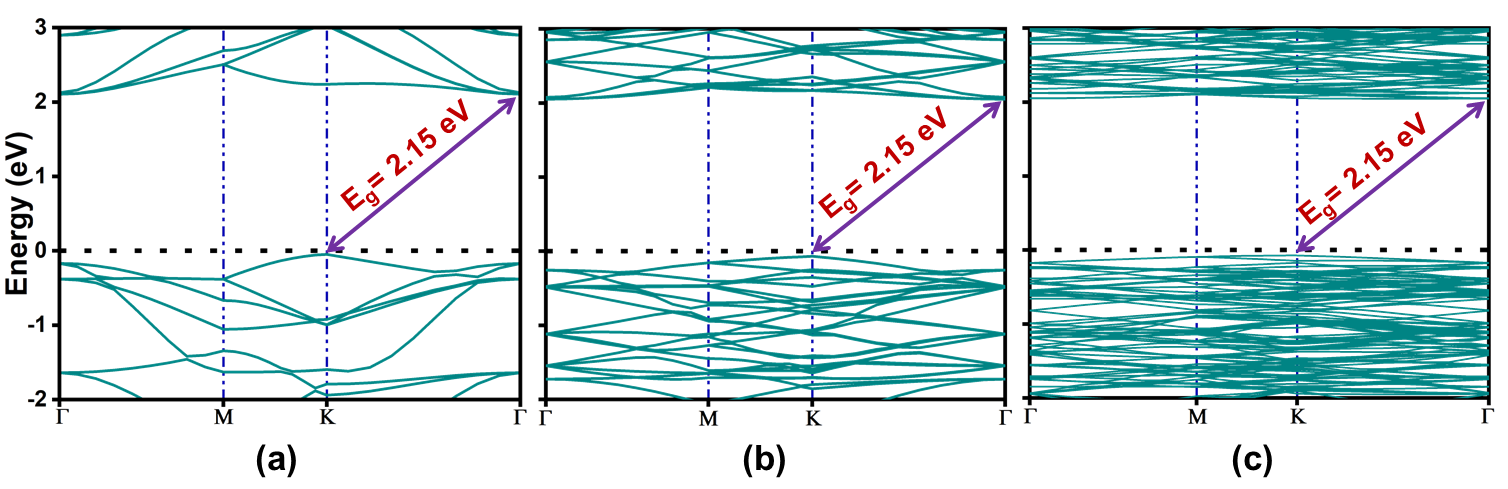}
\par\end{centering}
\caption{\label{fig:2}Calculated electronic band structures of (a) $2\times2$,
(b) $4\times4$, and (c) $8\times8$ GaP monolayers using GGA-PBE
functional. }

\end{figure}

\textcolor{black}{In order to check the convergence of the band structure
with respect to the supercell size, we computed the electronic band
structures of the $2\times2$, $4\times4$, and $8\times8$ supercells
of pristine GaP monolayer using GGA-PBE \citep{perdew1996} functional
as illustrated in Fig. \ref{fig:2}. From the figure it is obvious
that except for the total number of bands, the band structure for
the three supercells are virtually identical. In all the three cases,
the valence band maximum (VBM) and conduction band minimum (CBM) are
located at K and $\Gamma$ points, respectively, with an indirect
band gap of 2.15 eV, in agreement with the values reported in literature
for the $2\times2$ and $4\times4$ supercells \citep{suzuki2015,kumar2016}.
Since the band structures computed for the three supercells show a
similar trends, we have continued our investigation of pristine and
doped monolayers using the $4\times4$ supercell. }

\begin{figure}[H]
\begin{centering}
\includegraphics[scale=0.35]{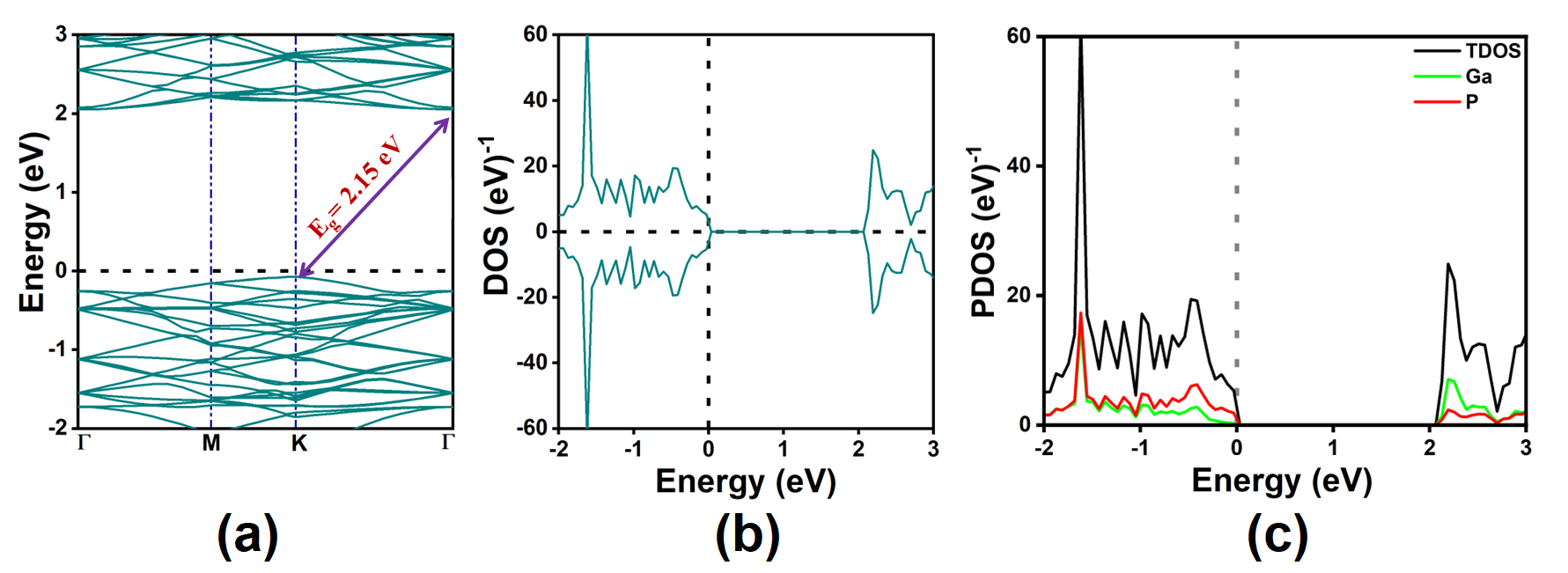}
\par\end{centering}
\caption{\label{fig:3}(a) Electronic band structure, (b) Density of states
(DOS), and (c) projected density of states (PDOS) of the pristine
GaP monolayer \textcolor{black}{using GGA-PBE \citep{perdew1996}
functional}. Black dotted line represents the Fermi level ($E_{F}$)
which is set at 0.00 eV.}
\end{figure}

\begin{figure}[H]
\begin{centering}
\includegraphics[scale=0.6]{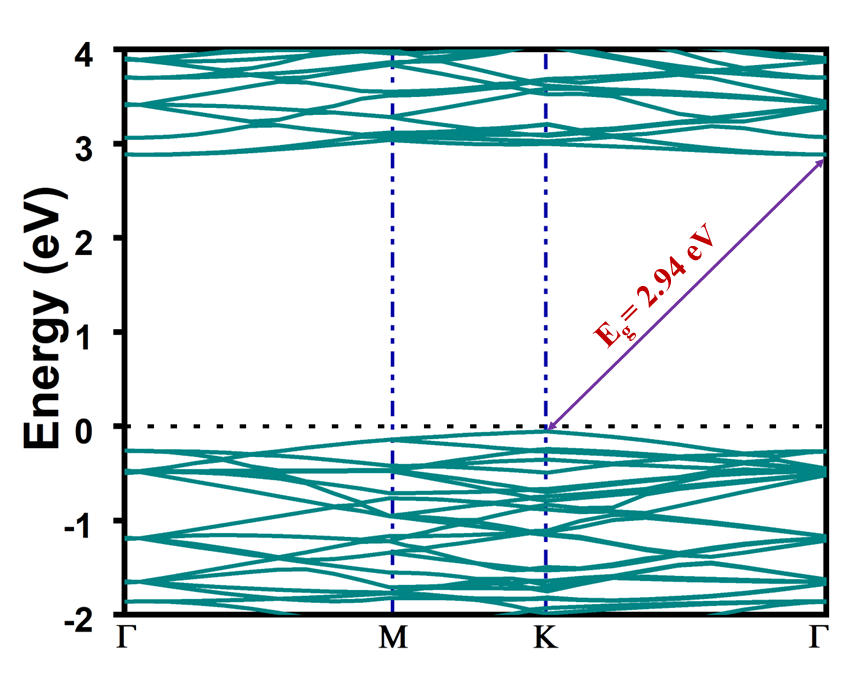}
\par\end{centering}
\caption{\label{fig:pris-hse} \textcolor{black}{Electronic band structure
of the pristine GaP monolayer using HSE06 functional. Black dotted
line represents the Fermi level ($E_{F}$) which is set at 0.00 eV.}}

\end{figure}

We present the band structure, density of states (DOS), and atom projected
density of states (PDOS) of the pristine $4\times4$ GaP monolayer,
computed using\textcolor{black}{{} GGA-PBE \citep{perdew1996} functional}
in Fig.\ref{fig:3}. To further probe the electronic structure, we
calculated the corresponding DOS (Fig. \ref{fig:3}(b)) which shows
good agreement with the band structure. Given the mirror symmetry
between up- and down-spins in the DOS plot, it is obvious that the
pristine GaP monolayer is non-magnetic in nature. To understand the
atomic contributions to the electronic structure of the pristine GaP
monolayer, we present its PDOS in Fig. \ref{fig:3}(c) from which
it is obvious that the valence band derives more contributions from
the P-based orbitals, while the conduction band is dominated by Ga
orbitals. \textcolor{black}{The band structure of the pristine $4\times4$
GaP monolayer is also calculated using hybrid exchange correlation
functional, HSE06 \citep{heyd2003hybrid}, and is presented in Fig.
\ref{fig:pris-hse}. The calculated band structure also predicts an
indirect band gap (K$\rightarrow\Gamma$) with the nature of energy
states identical to that of the GGA-PBE counterpart (Fig. \ref{fig:2}(a)).
However, as expected, the band gap gets increases to 2.94 eV which
shows excellent agreement with the previously reported value \citep{suzuki2015}.}

\subsubsection{Doped monolayers: Induced Magnetism}

Next, we investigate the influence of substitutional impurities on
the properties of the GaP monolayer. Recently, there have been several
studies of the induced magnetism in non-magnetic materials by impurity
doping in order to investigate their potential for magnetic applications
\citep{sethulakshmi2019,zheng2015}. Therefore, we also decided to
investigate the magnetic properties of the doped GaP monolayers by
performing spin-polarized calculations and computing the magnetic
moments, the results of which are presented in Tables \ref{tab2}
and \ref{tab3}. We note that there is a significant amount of nonzero
magnetic moment in the supercell when a single Ga is replaced by a
group IV element, or by transition metal (TM) atom Zn/Cd. In fact,
the maximum induced moment is for Sn doping. For the case of P substitution
(Table \ref{tab3}), we obtain finite magnetic moments for chalcogens
(except O) and TM doping, with the maximum moment obtained for Zn
doping.\textcolor{black}{{} To determine the nature of the induced magnetism,
we have performed spin polarized calculations corresponding to ferromagnetic
(FM) and antiferromagnetic (AFM) configurations for each such monolayer.
For these calculations, we have used the extended supercell of size
$8\times4\times1$, as discussed in the literature \citep{Hamid2022,Hamid2023}.
The energy difference between the FM and AFM states is calculated
as $\Delta E=E_{FM}-E_{AFM}$ and presented in Table \ref{tab:Calculated-energy-difference}.
According to this definition of $\Delta E$, its negative (positive)
value suggests the stability of the corresponding doped GaP monolayer
in the FM (AFM) state. Consequently, it is clear from the obtained
values of $\Delta E$ that Ga(C)P, Ga(Ge)P, and GaP(Cd) structures
are stable, in the FM state, whereas the other ones are in the AFM
state.}

\textcolor{blue}{}
\begin{table}[H]
\textcolor{black}{\caption{\label{tab:Calculated-energy-difference}\textcolor{black}{Calculated
energy difference ($\Delta E$) between the FM and AFM states for
the magnetism induced structures in case of both the Ga and P-substituted
monolayers.}}
}
\centering{}\textcolor{black}{}%
\begin{tabular}{ccccccc}
\toprule 
\multirow{2}{*}{\textcolor{black}{Systems}} & \multicolumn{6}{c}{\textcolor{black}{Ga-substituted monolayers}}\tabularnewline
\cmidrule{2-7} \cmidrule{3-7} \cmidrule{4-7} \cmidrule{5-7} \cmidrule{6-7} \cmidrule{7-7} 
 & \textcolor{black}{Ga(C)P} & \textcolor{black}{Ga(Si)P} & \textcolor{black}{Ga(Ge)P} & \textcolor{black}{Ga(Sn)P} & \textcolor{black}{Ga(Zn)P} & \textcolor{black}{Ga(Cd)P}\tabularnewline
\midrule 
\textcolor{black}{$\Delta E$ (eV)} & \textcolor{black}{-1.33} & \textcolor{black}{0.004} & \textcolor{black}{-0.004} & \textcolor{black}{0.003} & \textcolor{black}{0.004} & \textcolor{black}{0.005}\tabularnewline
\midrule
\midrule 
\textcolor{black}{Systems} & \multicolumn{6}{c}{\textcolor{black}{P-substituted monolayers}}\tabularnewline
\midrule 
 & \textcolor{black}{GaP(S)} & \textcolor{black}{GaP(Se)} & \textcolor{black}{GaP(Te)} & \textcolor{black}{GaP(Zn)} & \textcolor{black}{GaP(Cd)} & \tabularnewline
\midrule 
\textcolor{black}{$\Delta E$ (eV)} & \textcolor{black}{0.0004} & \textcolor{black}{0.0009} & \textcolor{black}{0.0007} & \textcolor{black}{0.013} & \textcolor{black}{-3.405} & \tabularnewline
\bottomrule
\end{tabular}
\end{table}

\begin{figure}[H]
\begin{centering}
\includegraphics[scale=0.5]{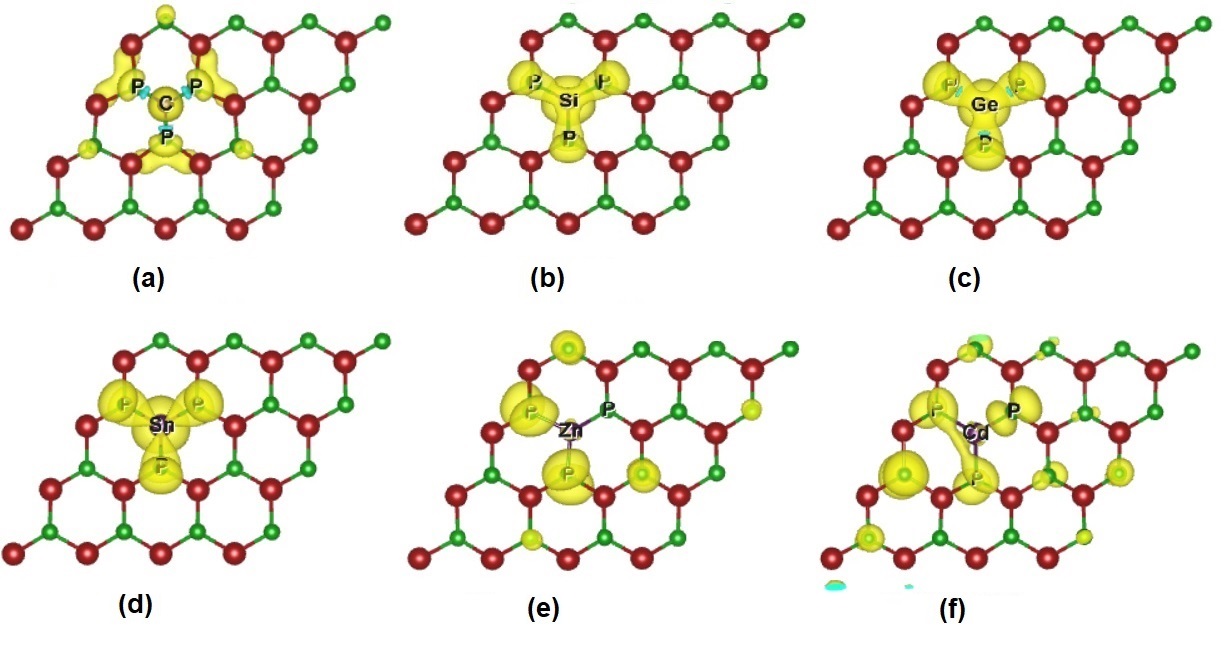}
\par\end{centering}
\caption{\label{fig:4}Spin density difference plots for Ga-substituted monolayers:
(a) Ga(C)P, (b) Ga(Si)P, (c) Ga(Ge)P, (d) Ga(Sn)P, (e) Ga(Zn)P, and
(f) Ga(Cd)P. Here yellow and blue colors represent the majority and
minority spin densities, respectively.}

\end{figure}

\begin{figure}[H]
\begin{centering}
\includegraphics[scale=0.5]{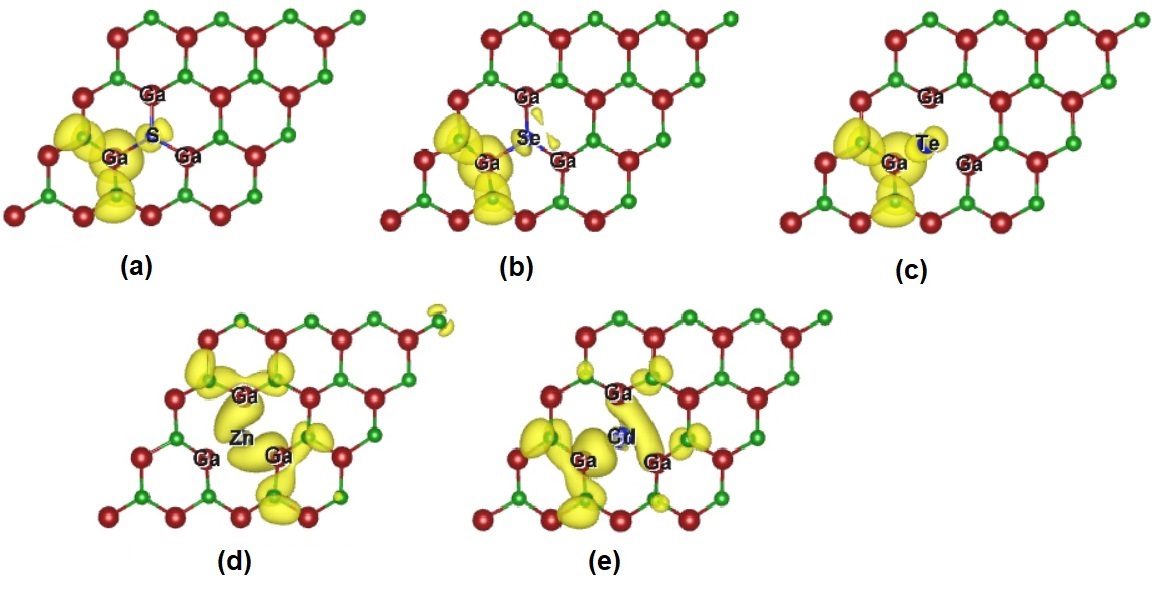}
\par\end{centering}
\caption{\label{fig:5}Spin density difference plots for P-substituted monolayers:
(a) GaP(S), (b) GaP(Se), (c) GaP(Te), (d) GaP(Zn), and (e) GaP(Cd).
Yellow and blue colors represent the majority and minority spin densities,
respectively.}

\end{figure}

To further study the induced magnetism, we have computed the spin
density differences, and plotted them in Figs. \ref{fig:4} and \ref{fig:5},
for the Ga- and P-substituted systems, respectively. For the Ga-substituted
systems , we note that except for the TM doped monolayers, major contribution
to the induced magnetism comes from the dopant and nearby atoms (Figs.
\ref{fig:4}(a),(b),(c),(d)). However, for the Zn and Cd doped monolayers,
their nearest neighbor elements are responsible for the induced magnetism
(Figs.\ref{fig:4}(e),(f)). From Fig.\ref{fig:5}, it is obvious that
for P-substituted systems, the dominant contribution to the magnetic
moment is from the atoms near to the doped site. 

Next, we will discuss the calculated spin-polarized electronic band
structures for these magnetic systems, and electronic band structures
for other impurities doped GaP monolayer which remains non-magnetic. 

\subsubsection{Band Structure of Ga-substituted monolayers}

\begin{figure}[H]
\begin{centering}
\includegraphics{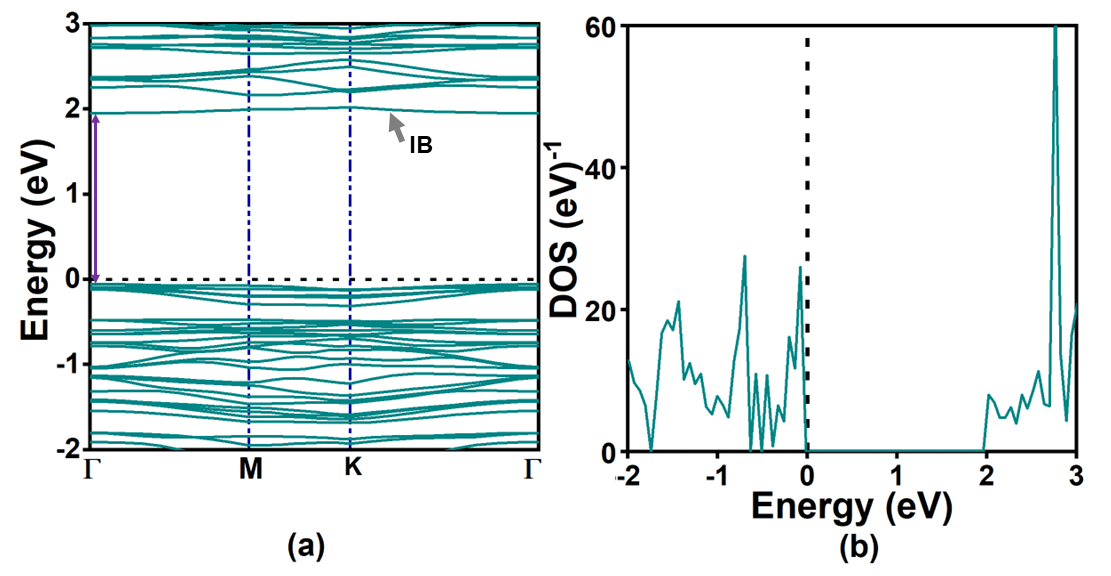}
\par\end{centering}
\caption{\label{fig:6}(a) Band structure, and (b) DOS plots for the Ga(In)P
structure. The impurity band (IB) is indicated by an arrow.}
\end{figure}

First, we discuss the band structures of substitutionally doped GaP
monolayers with one Ga atom of the supercell replaced by group III
elements, namely, B, Al and In. In Fig. \ref{fig:6} we present the
band structure and DOS of Ga(In)P, while in Fig. S3 of the SI, corresponding
plots for Ga(B)P and Ga(Al)P are presented. The band gaps for all
the monolayers are listed in Table \ref{tab2}. When compared to the
pristine case, we note that for all the three doped cases, a prominent
impurity band (IB) appears just below the lowest conduction band in
the gap region. This leads to a lowering of the electronic band gap
to 1.81 eV, 2.00 eV, and 2.00 eV for B, Al, and In dopants, respectively.
We consider this lowering to be significant in light of the fact that
the doping percentage is only 6.25. As far as the nature of the band
gap is concerned, it remains indirect for B- and Al-doped monolayers,
however, for the In-doped case, the band gap transforms quite interestingly
into a direct one at the $\Gamma$ point. This implies that in Ga(In)P,
optical transitions are possible across the gap, which clearly has
important technological implications for the design of GaP monolayer
based optoelectronic devices. \textcolor{black}{In addition to the
IB near the conduction band, several new bands appear near the top
of the valence band, thereby increasing the DOS in that region, which
is obvious from the DOS plots of Figs. \ref{fig:6} and S3. The PDOS
for the three doped monolayers is plotted in Fig. S4 of the SI. As
far as the contribution of Ga and P atoms to the PDOS is concerned,
the trends are unchanged as compared to the pristine monolayer, with
P contributing more to the valence bands, while Ga contributing more
to the conduction bands. As far as the impurity atoms are concerned,
predictably their contributions are small and concentrated near the
Fermi energy }$E_{F}$\textcolor{black}{.}\textcolor{blue}{{} }

\begin{figure}[H]
\begin{centering}
\includegraphics[scale=0.6]{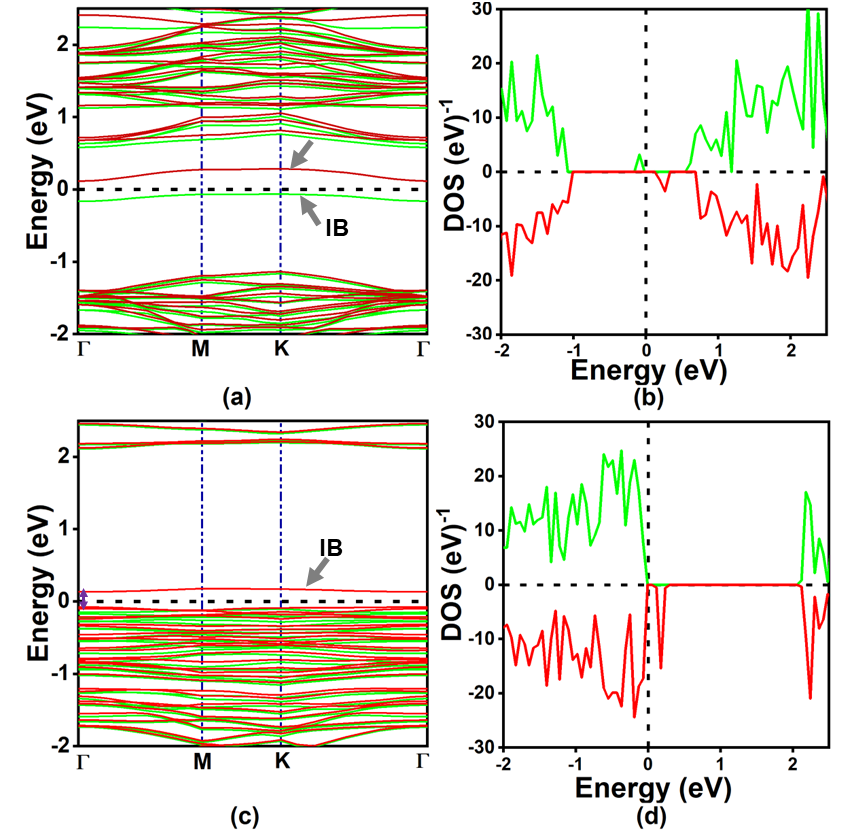}
\par\end{centering}
\caption{\label{fig:7} (a) and (b) band structure and DOS, respectively, for
Ga(C)P, (c) and (d) band structure and DOS, respectively, for Ga(Zn)P
structures. Green and red lines represent energy states corresponding
to majority and minority spin electrons, respectively. Impurity bands
(IB) are indicated by arrows.}
\end{figure}

The spin-polarized electronic band structures are calculated for those
Ga-substituted cases which resulted in nonzero magnetic moments discussed
in the previous section. In Fig.\ref{fig:7} we present the calculated
spin-polarized band structures for Ga(C)P and Ga(Zn)P systems from
which the splitting of energy states corresponding to charge carriers
of majority (green) and minority (red) spins is obvious. The nature
of band gap remains indirect corresponding to both the spin carriers
for Ga(C)P system (Fig.\ref{fig:7}(a)), whereas it becomes direct
for minority spin carriers when doped with Zn (Fig.\ref{fig:7}(c)).
This suggests the interesting possibility of spin-dependent optical
transitions across the direct bang gaps. It is worth mentioning that
the impurity doping resulted in additional energy states. However,
unlike the non-magnetic cases (B, Al, and In doping), large number
of additional energy states (ranging from 1.6 eV to 2 eV) get induced
in the forbidden energy region due to doping of C atom as it plays
a role of donor impurity. Also, these induced energy states corresponds
to the charge carriers of both the majority and minority spins. This
is also confirmed by the corresponding DOS presented in the same figure
(Fig.\ref{fig:7}(b)) which shows the emergence of peaks in this particular
region. Also, the asymmteric DOS profile corresponding to two spin
orientations verifies the induced magnetism in these doped monolayers.
Additionally, it is quite interesting to observe that one of the impurity
energy states which splits into majority and minority charge carriers
appeared near the Fermi energy level (indicated by arrow). Similar
trend is also observed in the electronic band structure of Ga(Si)P,
Ga(Ge)P, Ga(Sn)P, and Ga(Cd)P systems shown in Fig. S5 of the SI.
The corresponding DOS presented in the same figure shows good agreement
with their band structures. The magnitude as well as nature of the
resultant band gap are discussed in Table \ref{tab2}. Unlike other
magnetism induced cases, the band gap becomes direct for the charge
carriers of both spin orientations for the Ga(Cd)P system. Another
noteworthy point is the significant difference between the band gaps
of majority and minority spin carriers for the doped monolayers Ga(C)P,
Ga(Sn)P, Ga(Zn)P, and Ga(Cd)P. In particular for Ga(Zn)P, and Ga(Cd)P,
the minority carrier band gap almost closes, i.e., they are almost
half-metallic, suggesting their possible applications in spintronic
devices. From the PDOS shown in Fig. S6 of the SI, \textcolor{black}{we
can conclude that similar to the pristine}\textcolor{blue}{{} }monolayer,
the maximum contribution in valence band comes from the valence orbital
of P, whereas valence orbital of Ga contributes significantly to the
conduction band. \textcolor{black}{Dopant atom contributes to the
states near }$E_{F}$\textcolor{black}{{} where the impurity bands are
located.}\textcolor{blue}{{} }

\begin{figure}[H]
\begin{centering}
\textcolor{black}{\includegraphics{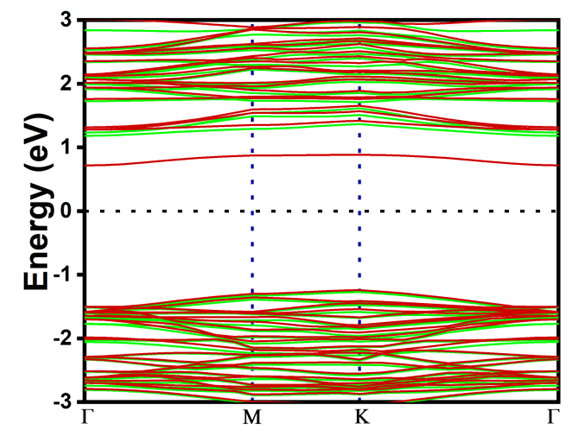}}
\par\end{centering}
\textcolor{black}{\caption{\textcolor{blue}{\label{fig:Ga-dop_hse}}\textcolor{black}{The electronic
band structure of the Ga(C)P structure calculated using the HSE06
functional. The green and red lines represent the bands corresponding
to majority and minority spin electrons, respectively. Black dotted
line represents the Fermi level ($E_{F}$) which is set at 0.00 eV.}}
}

\end{figure}

\textcolor{black}{The band structure is also calculated using the
HSE06 functional only for one of the Ga-replaced cases, i.e, Ga(C)P,
due to the high computational costs of such calculations, and the
results are presented in Fig. \ref{fig:Ga-dop_hse}. The HSE06 functional
increased the $E_{g}$ corresponding to both the majority and minority
spin carriers to 2.44 eV and 1.96 eV from the earlier values of 0.64
eV and 1.25 eV, respectively, obtained using PBE functional. The nature
of the bands remains the same and therefore the nature of the gap
remains indirect for both the spin carriers. }

\textcolor{black}{}

\textcolor{black}{Substitutional doping by different impurity elements
also leads to changes in the Fermi energy ($E_{F}$) as compared }to
that of the pristine GaP monolayer. It is observed from Table \ref{tab2},
that $E_{F}$ in case of doping with the same group elements shifts
towards the valence band thus showing a p-type behavior. Similar behavior
is observed for TM impurities (Zn/Cd) and the reason is that substitutional
doping by these elements in place of the Ga atom leads to their bond
formation with a P atom. But, Ga has three valence electrons whereas,
the TM atom contains only two valence electrons, thus resulting in
deficiency of one electron in forming a bond with P atom as compared
to the pristine Ga-P bond, leading to hole formation, i.e., p-type
doping. However, doping by group IV elements resulted in upward shift
(towards the conduction band) in $E_{F}$ because of one extra electron
than in their outer shell as compared to the Ga atom, thereby effectively
leading to n-type doping.

\subsubsection{Band Structure of P-substituted monolayers}

\begin{figure}[H]
\begin{centering}
\includegraphics[scale=0.6]{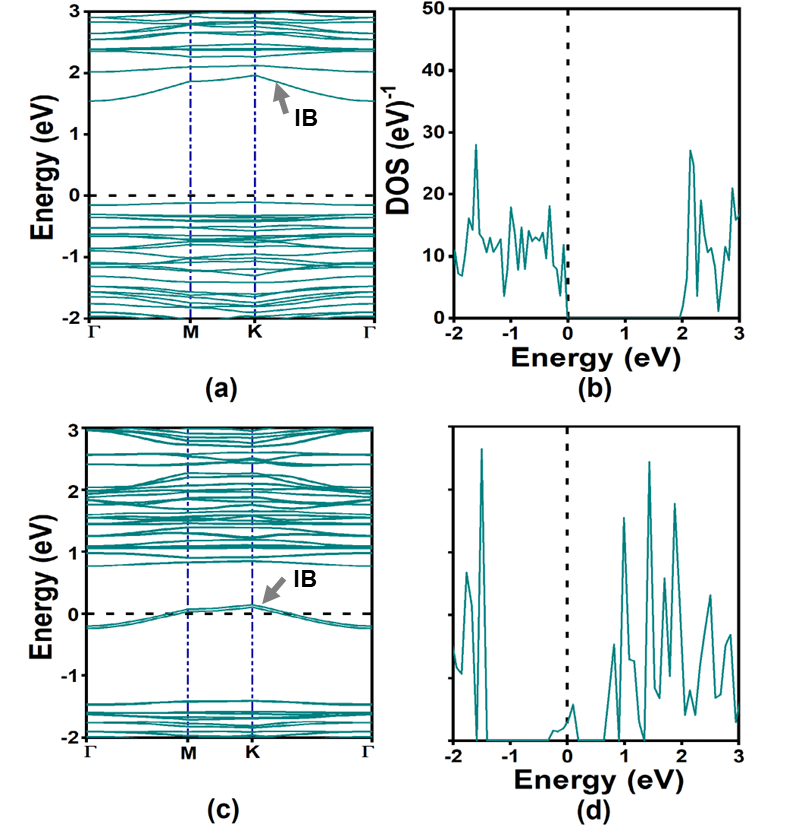}
\par\end{centering}
\caption{\label{fig:8} Electronic band structure/DOS plots for (a)/(b) GaP(N),
and (c)/(d) GaP(O) monolayers. The impurity bands (IB) are indicated
by arrows.}

\end{figure}

Next, we discuss the results of our calculations on the P-substituted
monolayers of the type GaP(Y). As discussed earlier, for Y=N, As,
Sb and O, our calculations predict non-magnetic ground states. The
calculated band structures of N and O-doped GaP monolayers are presented
in Fig. \ref{fig:8}, from which it is obvious that a single impurity
band gets formed around 1.5 eV in the gap region, for the N-doped
monolayer. Due to this, the magnitude of the band gap reduces significantly
to 1.64 eV, with still an indirect band gap. However, a drastic change
is observed for the GaP(O) system which turns metallic due to the
dopant O atom, which may be useful from a device perspective \citep{lu2007}.
All these modifications in the band structure after N and O doping
are validated by the corresponding DOS plots of Figs. \ref{fig:8}(b)
and \ref{fig:8}(d), respectively. The electronic band structure and
corresponding DOS for GaP(As) and GaP(Sb) structures are plotted in
Fig. S7 of the SI. The variations in the band gaps as a function of
doping elements are presented in Table \ref{tab3}. We note that the
As and Sb impurities changed the band gap of GaP monolayer very slightly
to 2.07 eV and 2.13 eV, respectively, while retaining the indirect
nature of the gap. \textcolor{black}{In Fig. S8 of the SI we plot
the PDOS for all the four dopants (N, As, Sb and O). We note that
the contributions of P and Ga atoms to PDOS near} $E_{F}$\textcolor{black}{{}
exhibit the same trends as for the pristine monolayer. However, the
dopant atom contributes to the PDOS near }$E_{F}$\textcolor{black}{{}
only for Sb doping. For other monolayers, it contributes to states
further away from $E_{F}$. For the O-doped monolayer, the states
at }$E_{F}$\textcolor{black}{{} leading to its metallic nature are
derived predominantly from the P and Ga atoms.}\textcolor{blue}{{} }

\begin{figure}[H]
\begin{centering}
\includegraphics[scale=0.6]{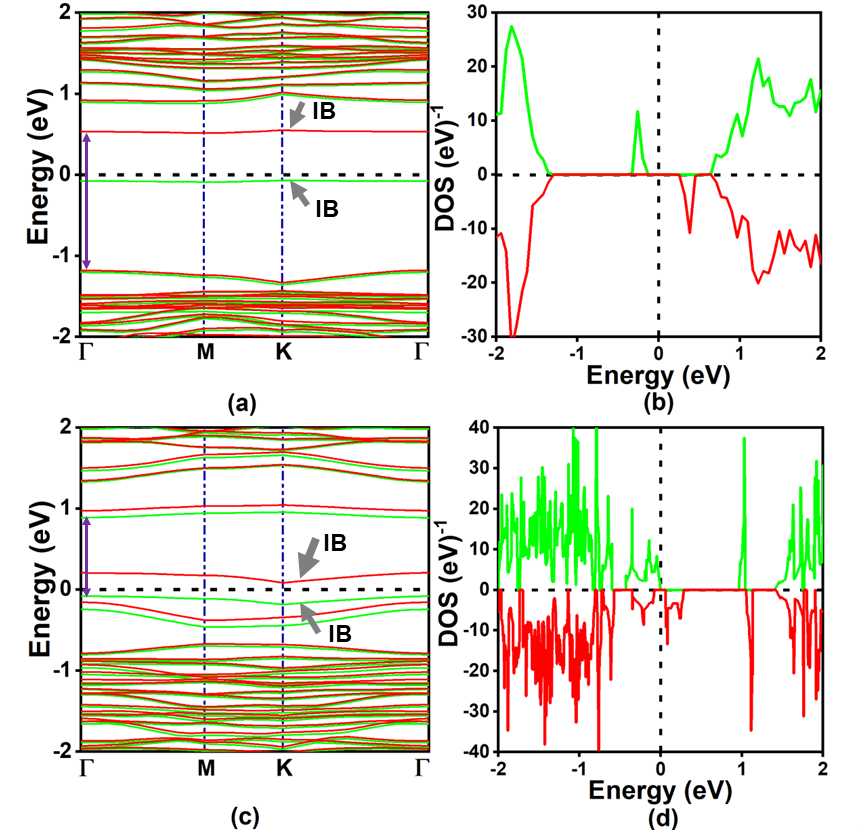}
\par\end{centering}
\caption{\label{fig:9}Electronic band structure/DOS for (a)/(b) GaP(S), and
(c)/(d) GaP(Zn) systems. Here green and red color lines represent
energy states corresponding to majority and minority spin charge carriers
respectively. Impurity bands (IB) are indiacted by arrow.}

\end{figure}

The spin-polarized electronic band structures are calculated for the
P-substituted GaP monolayer with chalcogens (S, Se, and Te) and TM
(Zn/Cd) as the dopants. The electronic band structures and corresponding
DOS for two representative monolayers GaP(S) and GaP(Zn) are presented
in Fig.\ref{fig:9}. We note that some additional bands, which are
quite flat, have appeared near the Fermi level corresponding to the
charge carriers of both the spins, denoted as IB in the figure. In
addition to the IBs close to $E_{F}$, some other IBs also emerge
which are somewhat farther from the Fermi level. Transition from the
indirect to a direct band gap is observed for minority (majority)
spin electrons for doping with S (Zn) atoms. The spin-polarized electronic
band structures and corresponding DOS for Se, Te, and Cd doped cases
are shown in Fig. S9 of the SI. The direct (indirect) band gap is
observed for the minority (majority) spin electrons both for Se and
Te doping, similar to the case of GaP(S) monolayer. Whereas, opposite
behavior is observed for Cd doping, with the majority band gap becoming
direct, \textcolor{black}{just as in the case}\textcolor{magenta}{{}
}of the GaP(Zn) layer. Again, the asymmetric nature of DOS in all
the cases corresponding to majority and minority spin electrons is
consistent with the induced magnetism. The magnitude of resultant
band gaps for all the cases are presented in Table \ref{tab3} from
which again a strong tendency towards half metallicity is obvious
for TM (Zn, Cd) doped GaP monolayers. The calculated PDOS (Fig. S10
of the SI) for induced magnetism cases shows that again the maximum
contribution in the valence and the conduction bands comes from the
valence orbitals of P and Ga atoms, respectively, with important contributions
by the dopant atoms to the impurity bands near $E_{F}$. 

\begin{figure}[H]
\begin{centering}
\includegraphics{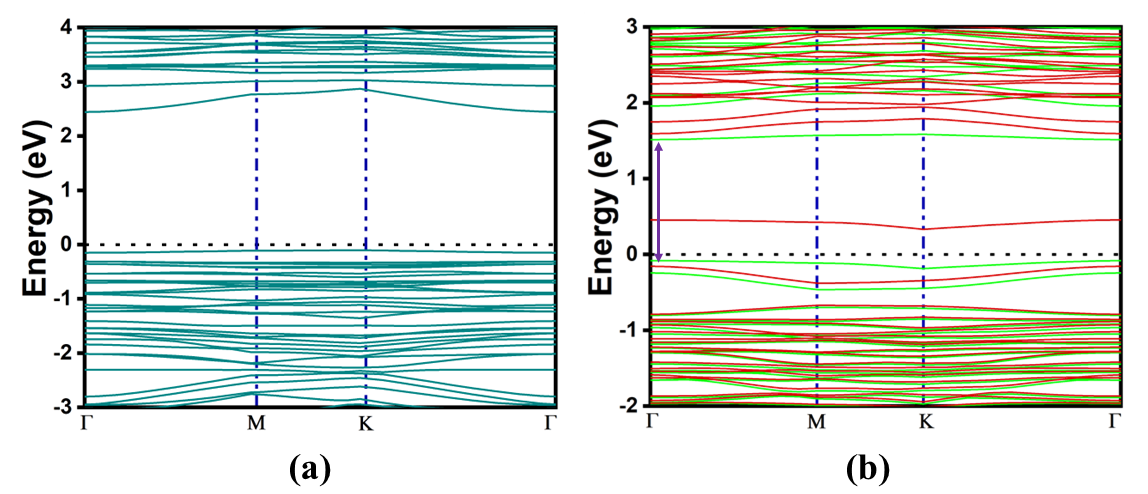}
\par\end{centering}
\caption{\textcolor{blue}{\label{fig:P-dop_hse}}\textcolor{black}{The electronic
band structures of the (a) GaP(N) and (b) GaP(Zn) monolayers calculated
using the HSE06 functional. The green and red lines represent the
bands corresponding to majority and minority spin electrons, respectively.
Black dotted line represents the Fermi level ($E_{F}$) which is set
at 0.00 eV.}}
\end{figure}

\textcolor{black}{Next, we present the HSE06 band structures for the
two P-replaced cases, i.e., GaP(N) and GaP(Zn), in Fig. \ref{fig:P-dop_hse},
and note that while the GaP(N) monolayer remains nonmagnetic, the
GaP(Zn) monolayer becomes magnetic, in agreement with the PBE results.
Expectedly, the band gaps get increased to 2.55 eV for the GaP(N)
system, and to 1.6 eV (0.48 eV) corresponding to majority (minority)
spin carriers of the GaP(Zn) system. However, the nature of the bands
and the band gaps remain identical to those obtained using PBE functional,
i.e. indirect for GaP(N) system, whereas, direct (indirect) for GaP(Zn)
system corresponding to majority (minority) spins.}

The variation in $E_{F}$ as a function of dopant elements, as presented
in Table \ref{tab3}, also shows interesting behavior. The doping
of N, Sb, O, and Te impurities shifted the $E_{F}$ towards the valence
band leading to effectively p-type doping as compared to the pristine
counterpart, whereas for S, Se, Zn and Cd doping, it shifted towards
the conduction band leading to n-type doping. However, the $E_{F}$
remains unchanged in case of As doped GaP monolayer. 

\subsubsection{Bader charge analysis}

To investigate the influence of doping on the charge distribution
in the monolayers, we have performed the Bader charge analysis, and
the results are presented in Table \ref{tab:4}. In the table we present
the Bader charges on the doping site (Ga/P), and its three nearest-neighbor
atoms (P/Ga) before and after the doping. We note that in all the
cases of doping at the Ga site (except for C doping), the charge gets
transfered from the dopant to the nearest neighboring P atoms due
to the relatively higher electronegativity of P atoms than the dopants.
However, the opposite result is observed for C doping because P atom
is less electronegative as compared to the C atom. For the cases of
P-site substitution, it is observed that the dopants (except Zn and
Cd) gain charge from its nearest atoms (Ga), the reason for which
can again be explained in terms of electronegativity differences,
i.e., dopants are more electronegative than the Ga atom. In case of
TM doping, we get an interesting result that both the dopant and Ga
atoms lose charge because of their comparable electronegativities.
As a result, the charges from the dopant (Zn and Cd) and Ga atoms
get transferred to the nearest P atoms, which are more electronegative
as compared to them. 
\begin{table}[H]
\caption{\label{tab:4}Charge accumulated on dopant and its nearest neighbor
atoms for both the Ga and P substituted structures, i.e. Ga(X)P and
GaP(Y).}

\centering{}%
\begin{tabular}{cccc|cccc}
\hline 
\multicolumn{8}{c}{Accumulated charge (e)}\tabularnewline
\hline 
\multicolumn{4}{c|}{Ga(X)P} & \multicolumn{4}{c}{GaP(Y)}\tabularnewline
\hline 
Dopant & P1 & P2 & P3 & Dopant & Ga1 & Ga2 & Ga3\tabularnewline
\hline 
Ga=2.26 & 5.73 & 5.73 & 5.73 & P=5.73 & 2.26 & 2.26 & 2.26\tabularnewline
\hline 
B=2.54 & 5.67 & 5.67 & 5.66 & N=5.98 & 2.32 & 2.32 & 2.32\tabularnewline
Al=1.10 & 6.12 & 6.12 & 6.12 & As=5.49 & 2.32 & 2.32 & 2.32\tabularnewline
In=2.29 & 5.69 & 5.69 & 5.69 & Sb=5.25 & 2.41 & 2.41 & 2.41\tabularnewline
C=5.35 & 4.98 & 4.98 & 4.96 & O=7.11 & 2.11 & 2.11 & 2.11\tabularnewline
Si=2.84 & 5.85 & 5.85 & 5.84 & S=6.82 & 2.44 & 2.17 & 2.17\tabularnewline
Ge=3.46 & 5.66 & 5.66 & 5.66 & Se=6.66 & 2.46 & 2.23 & 2.23\tabularnewline
Sn=3.30 & 5.70 & 5.70 & 5.70 & Te=6.39 & 2.50 & 2.32 & 2.32\tabularnewline
Zn=11.41 & 5.68 & 5.68 & 5.70 & Zn=11.89 & 2.54 & 2.54 & 2.51\tabularnewline
Cd=11.44 & 5.68 & 5.68 & 5.65 & Cd=11.90 & 2.50 & 2.50 & 2.50\tabularnewline
\hline 
\end{tabular}
\end{table}

\subsection{Optical Properties}

We saw in the previous section that for several cases of doping band
gap becomes direct, implying that those systems could be useful in
optoelectronic devices. Also in some cases a direct band gap is only
observed for electron's one spin orientation, implying spin-dependent
optoelectronic properties. Therefore, we decided to compute optical
absorption spectra of the studied monolayers, both for the pristine
and doped ones. For the purpose, we have calculated the dielectric
functions, refractive indices, and absorption coefficients for the
layers with a direct band gap after doping. The information regarding
the response of a material to an incident external electromagnetic
radiation of frequency $\omega$ is contained in the dielectric function
$\varepsilon(\omega)=\varepsilon_{1}(\omega)+i\varepsilon_{2}(\omega)$,
where $\varepsilon_{1}(\omega)$ and $\varepsilon_{2}(\omega)$ are
its real and imaginary parts, respectively. \textcolor{black}{The
imaginary part of dielectric function is calculated by taking summation
of the conduction band states, while its real part is obtained by
applying the Kramers-Kronig relationship.}\textcolor{blue}{{} }\textcolor{black}{Our
calculations of the optical absorption spectra were within the framework
of the independent particle model (IPM) in which only band-to-band
electronic transitions are considered. That is, the influence of electron
(e) -hole (h) interactions was neglected which give rise to the formation
of midgap excitons. To incorporate the e-h interactions, one needs
to employ methodologies like Bethe-Salpeter equation (BSE) \citep{BSE1954,waheed2023,waheed2023janus,amin2023,KHURAMI2023},
which go beyond IPM and include quantum many-body effects. Nonetheless,
the main objective of these calculations is to attain a qualitative
grasp of how defects change the optical properties of the GaP monolayer,
for which we feel that IPM suffices. However, for a quantitatively
accurate description of the optical properties of the GaP monolayer,
computationally demanding BSE approach will be necessary. The }refractive
index $n(\omega)$, and absorption coefficient $\alpha(\omega)$ are
also calculated using the equations

\begin{equation}
n(\omega)=\frac{1}{\sqrt{2}}\left[\sqrt{\epsilon_{1}^{2}(\omega)+\epsilon_{2}^{2}(\omega)}+\epsilon_{1}(\omega)\right]^{\frac{1}{2}}\label{eq:n-omega}
\end{equation}

\begin{equation}
\alpha(\omega)=\frac{\omega\sqrt{2}}{c}\left[\sqrt{\epsilon_{1}^{2}(\omega)+\epsilon_{2}^{2}(\omega)}-\epsilon_{1}(\omega)\right]^{\frac{1}{2}}\label{eq:alpha-omega}
\end{equation}

\begin{figure}[H]
\begin{centering}
\includegraphics[scale=0.5]{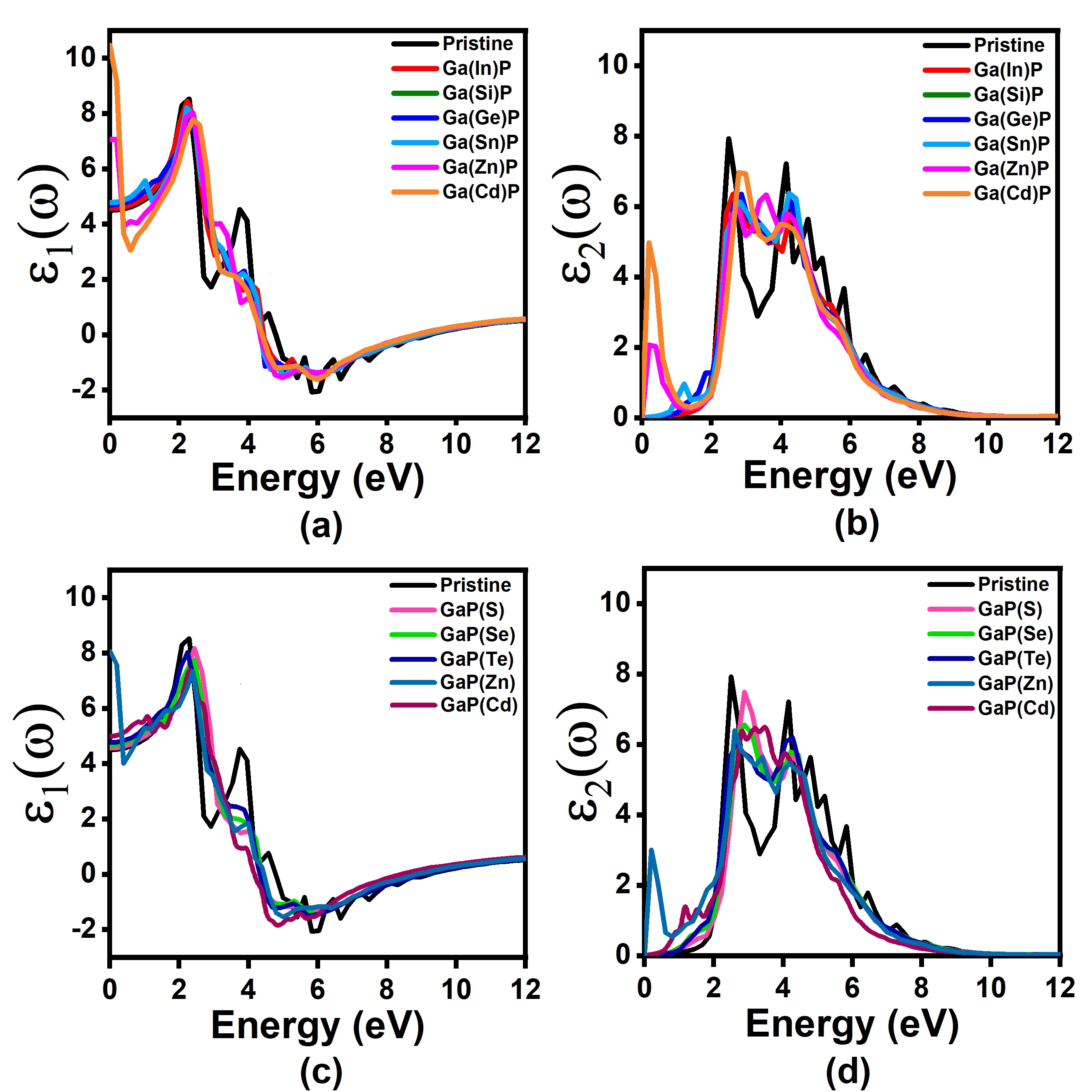}
\par\end{centering}
\caption{\label{fig:dielectric} Variations of real ((a) and (c)) and imaginary
parts ((b) and (d)) of the complex dielectric functions, respectively
for the Ga(X)P and GaP(Y) systems, with respect to the energy of the
incident light. }

\end{figure}

The real and imaginary parts of the dielectric function are plotted
for both Ga(X)P (Ga substitution) and GaP(Y) (P substitution) structures
in Fig. \ref{fig:dielectric}. The imaginary part of the dielectric
function experiences a red shift after doping in all the considered
cases (see Figs. \ref{fig:dielectric}(b),(d)) due to the doping-induced
band gap reduction. The real part of the dielectric function increases
for the doped monolayers as compared to the pristine one (see Figs.
\ref{fig:dielectric}(a),(c)). The high value of dielectric constant
reduces the amount of radiative recombination, consequently enhances
the device performance as far as solar cells are concerned \citep{deb2020,jain2020}. 

The refractive indices $n(\omega)$ calculated using Eq. \ref{eq:n-omega}
are presented in Fig. S11 of SI. Further, we have also calculated
the optical absorption coefficients $\alpha(\omega)$ (using Eq. \ref{eq:alpha-omega})
for the pristine and those doped GaP monolayers with direct band gaps
for at least one spin orientation, and compared them to that of the
pristine monolayer. 

\begin{figure}[H]

\begin{centering}
\includegraphics[width=15cm]{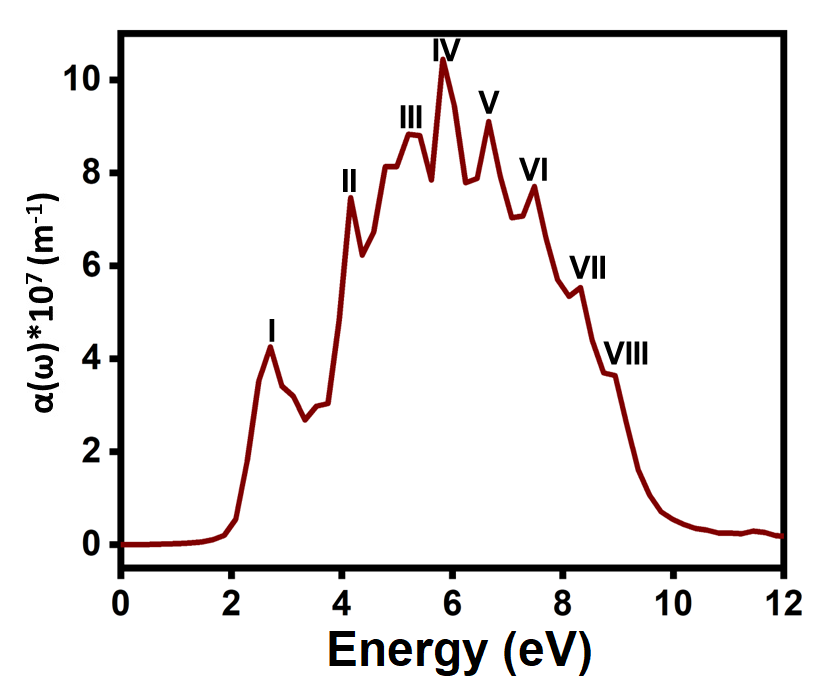}
\par\end{centering}
\caption{\label{fig:alpha-pristine}Optical absorption coefficient $\alpha(\omega)$
of the pristine GaP monolayer as a function of the energy of the incident
photon. }
\end{figure}

In Fig. \ref{fig:alpha-pristine} we plot the absorption coefficient
($\alpha(\omega)$) of the pristine GaP monolayer, which is fairly
broad spanning the energy region 2.0 eV -- 10.0 eV. Nevertheless,
the spectrum has several well-defined peaks which are labeled in the
figure, and whose positions are listed in Table \ref{tab:peaks-pristine}.\textcolor{magenta}{{}
}\textcolor{black}{Because the pristine monolayer is an indirect band
gap material,} the peaks in its absorption spectrum are located at
energies higher than the band gap of 2.15 eV. The first peak occurs
at 2.70 eV corresponding to the transition $v\rightarrow c+5$, at
the K point, where $v/c$ denote the highest/lowest valence/conduction
band. Similarly, the maximum intensity peak IV, located at 5.82 eV
is due to the transition $v\rightarrow c+34$, at the $\Gamma$ point.
The full list of transitions corresponding to various peaks of the
absorption spectra is presented in Table S2 of the SI from which it
is obvious that the absorption peaks at successively higher energies
involve transitions from the highest valence band into the high-lying
conduction bands. However, noteworthy point is that the spectra are
calculated within the mean-field approximation of DFT, and, thus do
not include the excitonic effects caused by the electron-hole interactions.
Nevertheless, by comparing the absorption spectrum of the pristine
monolayer to those of doped monolayers, we can achieve a qualitative
understanding of the influence of doping on the optical absorption. 

\begin{table}[H]
\caption{\label{tab:peaks-pristine}Excitation energies of various peaks appearing
in the optical absorption coefficient ($\alpha(\omega))$ of the pristine
GaP monolayer plotted in Fig. \ref{fig:alpha-pristine}. MI denotes
the most-intense peak.}

\centering{}%
\begin{tabular}{c|cccccccc}
\hline 
Peak & I & II & III & IV  & V & VI & VII & VIII\tabularnewline
\hline 
Energy (eV) & 2.70 & 4.16 & 5.20 & 5.82(MI) & 6.65 & 7.49 & 8.32 & 8.94\tabularnewline
\hline 
\end{tabular}
\end{table}

\begin{figure}[H]
\begin{centering}
\includegraphics[scale=0.5]{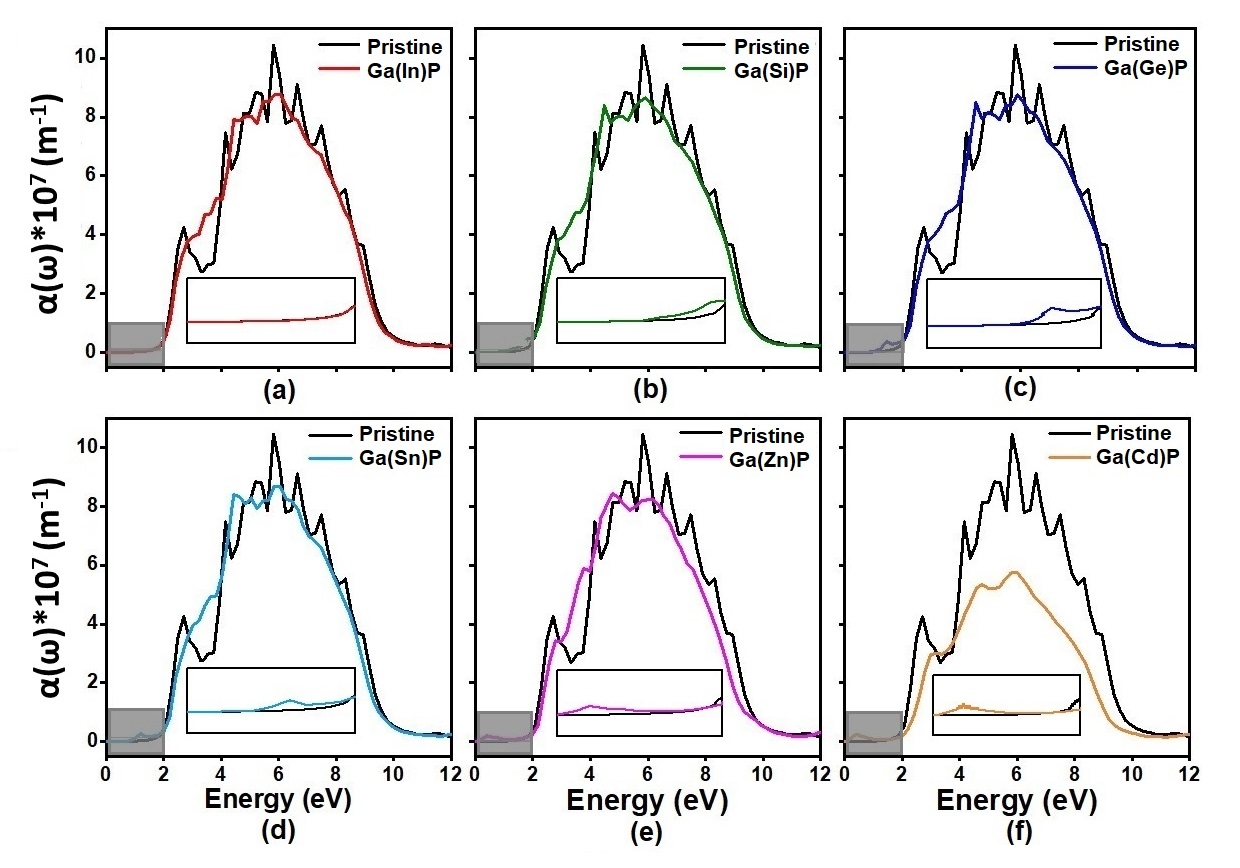}
\par\end{centering}
\caption{\label{fig:alpha-ga-sub-x}Optical absorption coefficient ($\alpha(\omega)$)
as a function of the energy of the incident photon for Ga-substituted
monolayers: (a) Ga(In)P, (b) Ga(Si)P, (c) Ga(Ge)P, (d) Ga(Se)P, (e)
Ga(Zn)P, and (f) Ga(Cd)P. Shaded regions are shown in inset.}

\end{figure}

\begin{figure}[H]
\begin{centering}
\includegraphics[scale=0.5]{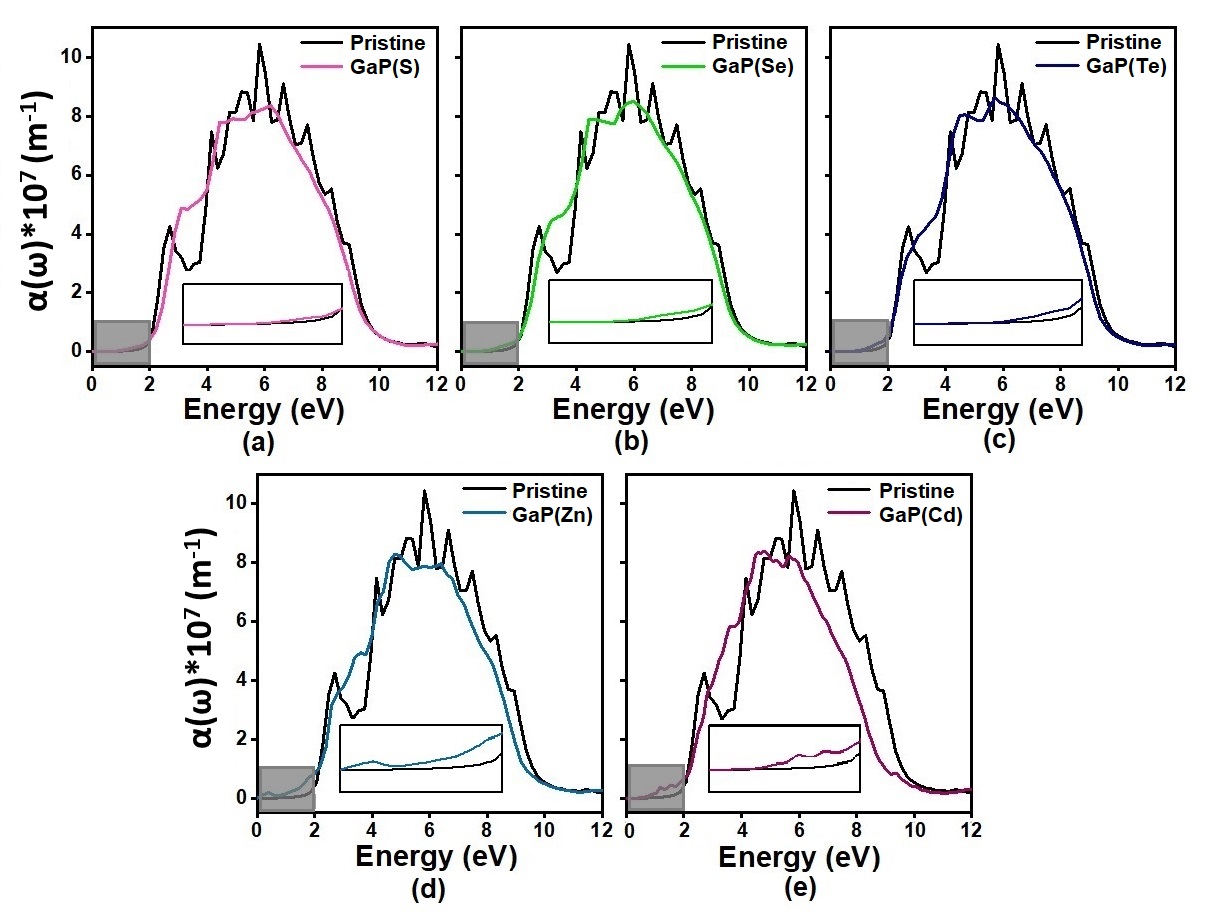}
\par\end{centering}
\caption{\label{fig:alpha-p-sub-x}Optical absorption coefficient ($\alpha(\omega)$)
as a function of the energy of the incident photon for P-substituted
monolayers: (a) GaP(S), (b) GaP(Se), (c) GaP(Te), (d) GaP(Zn), and
(e) GaP(Cd). Shaded regions are shown in inset.}
\end{figure}

The optical absorption coefficients plotted with respect to the energy
of the incident photon are presented in Figs. \ref{fig:alpha-ga-sub-x}
and \ref{fig:alpha-p-sub-x} for Ga(X)P and GaP(Y) systems, respectively.
The shaded region of each plot is shown in the inset from which it
is clear that in several cases, low-intensity \textcolor{black}{peaks}
emerge in the lower-energy regions of the spectra. To further illustrate
the changes in the absorption spectra due to doping, in Table \ref{tab:1-mi}
we present locations of the first ($E_{I}$) and the most-intense
peaks ($E_{MI}$) both for the pristine and the doped monolayers.
The changes in $E_{I}$ are clearly due to the impurity bands located
in the gap region of the pristine monolayer and can be correlated
with the modified band gaps of the doped monolayers presented in the
Tables \ref{tab2} and \ref{tab3}. For example, $E_{I}=0.40$ eV
for Ga(Zn)P can be understood in terms of its direct band gap of 0.20
eV for the down-spin channel (see Table \ref{tab2}). On the other
hand, changes in $E_{MI}$ are due to global changes in the band structure
caused by the dopant atoms.

\begin{table}[H]
\caption{\label{tab:1-mi}First peak position ($E_{I}$) and most-intense peak
($E_{MI}$) in terms of energy for both Ga(X)P and GaP(Y) systems.}

\centering{}%
\begin{tabular}{ccc|ccc}
\hline 
Ga(X)P  & $E_{I}$(eV) & $E_{MI}$(eV) & \multirow{1}{*}{GaP(Y)} & $E_{I}$(eV) & $E_{MI}$(eV)\tabularnewline
\hline 
Pristine  & \multirow{1}{*}{2.70} & 5.82 & Pristine & 2.70 & 5.82\tabularnewline
Ga(In)P & 3.23 & 6.03 & GaP(S) & 4.87 & 6.22\tabularnewline
Ga(Si)P & 1.84 & 5.92 & GaP(Se) & 1.77 & 6.00\tabularnewline
Ga(Ge)P & 1.43 & 5.9 & GaP(Te) & 1.83 & 5.70\tabularnewline
Ga(Sn)P & 1.21 & 5.86 & GaP(Zn) & 0.40 & 4.80\tabularnewline
Ga(Zn)P & 0.40 & 4.77 & GaP(Cd) & 1.15 & 4.81\tabularnewline
Ga(Cd)P & 0.40 & 5.97 &  &  & \tabularnewline
\hline 
\end{tabular}
\end{table}

From Figs. \ref{fig:alpha-ga-sub-x} and \ref{fig:alpha-p-sub-x},
it is clear that the optical absorption spectra of both the pristine
and doped monolayers cover a wide range of visible and ultraviolet
regions making them potential candidates in optoelectronic devices
and photovoltaic solar cells. Another interesting aspect of several
doped monolayers is the presence of spin-polarized bands with unequal
band gaps for the two spin orientations (see Table \ref{tab2}). This
suggests the possibility of spin-dependent optical response in these
materials which can also be used for device applications.\textcolor{red}{{} }

\textcolor{black}{We have also investigated the above said properties
(electronic and optical) by varying the doping percentage from 6.25
\% to 2.78 \% by using a $6\times6$ supercell, both for the pristine
and the doped GaP monolayers. This is done to study the effect of
variation of the doping concentration on the said properties. The
results of these calculations are discussed in the last section of
the SI.}

\section{Conclusion}

\label{sec:Conclusion}

To summarize, using a first-principles DFT-based methodology employing
the GGA-PBE exchange-correlation functional, we have computationally
studied the structural, electronic, magnetic, and optical properties
of both pristine and doped GaP monolayers. We find that the pristine
monolayer, instead of being completely flat, is periodically buckled
with an indirect band gap of 2.15 eV. We considered substitutional
doping of the pristine monolayer with a single Ga or P atom of the
supercell replaced by the impurity atom, and find that for all doping
configurations the resultant monolayers are stable. In several cases
the doping led to significant tuning of the band gap, \textcolor{black}{and
in one case} (In replacing Ga) the band gap transformed into a direct
one. Also, doping of some of the impurity elements resulted in induced
magnetism due to which the degeneracy of \textcolor{black}{spin up
and spin down} electrons got lifted, resulting in a spin-polarized
band structure. In some of these cases we noted that: (a) band gap
of one of the spin orientations was quite small indicating a tendency
towards half-metallic behavior, and (b) the band gap corresponding
to one of the spins became direct.\textcolor{magenta}{{} } The former
property can be utilized in spintronics while the latter one can be
exploited for spin-dependent optics. Thus, we also investigated the
optical properties for those doped structures in which for one or
both spin orientations the band gap was direct. The obtained optical
absorption spectrum spans over wide range of visible and ultraviolet
region which suggest its possible applications in solar cells and
other optoelectronic devices. Therefore, we believe that doped GaP
monolayers can be useful in a wide variety of device applications.
Furthermore, given the fact that monolayers have a much larger surface
area available as compared to their bulk counterparts, they can also
be studied for their catalytic performance. Research along those lines
is presently ongoing in our group, and the results will be conveyed
in future publications. We also hope that our work will stimulate
the experimental activity aimed at synthesizing pristine and doped
GaP monolayers.

\section*{acknowledgments}

One of the authors, K.D. acknowledges financial assistance from Prime
Minister Research Fellowship (PMRF). R.Y. acknowledges the support
through the Institute Post-Doctoral Fellowship (IPDF) of Indian Institute
of Technology Bombay. 

\bibliographystyle{iopart-num}
\nocite{*}
\bibliography{Reference-GaP}

\end{document}